\documentclass[11pt]{article}

\usepackage{url,cite}

\usepackage[latin1]{inputenc}
\usepackage[english]{babel}

\usepackage{amssymb,amsmath,amsthm,cancel,hyperref,graphicx,xcolor}
\usepackage{picinpar,graphicx,xypic}
\usepackage{booktabs}
\usepackage{mathrsfs}
\usepackage[font=small,labelfont=bf,margin=3mm,labelsep=period,tableposition=top]{caption}
\usepackage[a4paper, top=2.2cm, bottom=2.2cm, left=2cm, right=2cm, bindingoffset=0mm]{geometry}

\numberwithin{equation}{section}
\pdfoutput=1

\newcommand{\gsim}{\gtrsim}
\newcommand{\lsim}{\lesssim}
\newcommand{\mh}{m_H}
\newcommand{\mt}{m_t}
\newcommand{\mz}{m_Z}
\newcommand{\kt}{k_{\rm t}}
\newcommand{\MSbar}{\overline{\text{MS}}}
\newcommand{\sss}{\scriptscriptstyle\rm}
\renewcommand{\Re}{\mathrm{Re}\:}

\newcommand{\Ord}{\mathcal{O}}
\newcommand{\Lum}{\mathscr{L}}
\newcommand{\Mell}{\mathcal{M}}
\newcommand{\as}{\alpha_s}
\newcommand{\muf}{\mu_{\sss F}}
\newcommand{\mur}{\mu_{\sss R}}
\newcommand{\Rgg}{\mathcal{R}_{gg}}
\newcommand{\lf} {\ell_{\sss F}}

\newcommand{\plus}[1]{\left(#1\right)_+}
\newcommand{\plusq}[1]{\left[#1\right]_+}
\newcommand{\D}{\mathcal{D}}

\newcommand{\Dh}{\hat{\mathcal{D}}}
\newcommand{\Dm}{\mathcal{D}^{\log}}

\newcommand{\gammae}{\gamma_{\scriptscriptstyle E}}

\let\originalleft\left
\let\originalright\right
\renewcommand{\left}{\mathopen{}\mathclose\bgroup\originalleft}
\renewcommand{\right}{\aftergroup\egroup\originalright}

\newcommand{\Csoft}{C_{\text{soft}}}
\newcommand{\Csoftm}[1]{C_{\text{soft}_#1}}
\newcommand{\Chard}{C_{\text{h.e.}}}
\newcommand{\Cmv}  {C_{\text{$N$-soft}}}
\newcommand{\Cabf} {C_{\text{ABF}}}
\newcommand{\Csub} {C_{\text{ABF-sub}}}

\def\beq{\begin{equation}}  
\def\eeq{\end{equation}}
\def\({\left(}
\def\){\right)}
\def\[{\left[}
\def\]{\right]}

\graphicspath{{images/}}

\begin{document}
\begin{flushright}
DCPT/13/30\\  IPPP/13/15\\
DESY 13-001\\
Edinburgh 2012/25\\
IFUM-1010-FT\\
\end{flushright}

\begin{center}
{\Large \bf
Higgs production in gluon fusion beyond NNLO}
\vspace*{1.5cm}

Richard D.~Ball$^{a}$, Marco Bonvini$^b$, Stefano Forte$^{c}$, Simone Marzani$^d$ \\and Giovanni Ridolfi$^e$
\\
\vspace{0.3cm}  {\it
{}$^a$Tait Institute, University of Edinburgh,\\  Edinburgh EH9 3JZ, Scotland\\ \medskip
{}$^b$ Deutsches Elektronen-Synchroton, DESY,\\ Notkestra{\ss}e 85, D-22603 Hamburg, Germany\\ \medskip
{}$^c$Dipartimento di Fisica, Universit\`a di Milano and
INFN, Sezione di Milano,\\
Via Celoria 16, I-20133 Milano, Italy\\ \medskip
{}$^d$Institute for Particle Physics Phenomenology, Durham University,\\ Durham DH1 3LE, England\\ \medskip
{}$^e$Dipartimento di Fisica, Universit\`a di  Genova and INFN,
Sezione di Genova, \\Via Dodecaneso 33, I-16146 Genova, Italy
}
\vspace*{1.5cm}

\bigskip
\bigskip

{\bf \large Abstract:}
\end{center}

We construct an approximate expression for the cross section
for Higgs production in gluon fusion at
next-to-next-to-next-to-leading order (N$^3$LO) in $\as$ with finite top
mass. We argue that an accurate approximation 
 can be constructed by exploiting the analiticity of the
Mellin space cross section, and the information on its singularity
structure coming from large $N$  (soft gluon, Sudakov)
and small $N$ (high energy, BFKL) all order resummations. We support our argument
with an explicit comparison of the approximate and the exact expressions
up to the highest (NNLO) order at which the latter are available. We
find that the approximate N$^3$LO result amounts to a correction of
$16\%$ to the NNLO QCD cross section for production of a 125~GeV Higgs
at the LHC (8~TeV), larger than previously
estimated, and it significantly reduces the scale dependence of the NNLO result.

\clearpage

\section{Introduction} \label{sec:intro}

The dominant Higgs production mechanism at the LHC is gluon
fusion via a heavy fermion loop (mainly a top quark)~\cite{georgi},
and indeed the recent announcement of the discovery of a Higgs-like
particle~\cite{atlashiggs,cmshiggs} is largely based on events in this
channel. In view of this, an accurate
determination of the cross section in this channel is of great
interest. Next-to-leading order (NLO) corrections to the
inclusive cross section, originally computed in
Refs.~\cite{spira1,dawson} in the large top mass ($\mt\to\infty$)
approximation, and in Ref.~\cite{spira2} for general $\mt$ are known
to be as large as the leading order, and the NNLO corrections (first
computed in Refs.~\cite{harlanderNNLO,anastasiouNNLO, ravindranNNLO}
in the $\mt\to\infty$ limit and for finite top mass in
Refs.~\cite{Higgsfinite, harlander1,harlander2, harlander3, pak1, pak2})
about half as large as the leading order.
The significant scale dependence of the NNLO result suggests that
corrections at yet higher orders are not negligible: in fact they
currently account for half or more of the uncertainty on the theory
prediction for the cross section~\cite{Dittmaier:2011ti} (the other
half being due to parton distributions and the strong coupling).

While computations of the full N$^3$LO correction to the
cross section are in progress~\cite{anastasiouintegrals,Hoschele:2012xc,Anastasiou:2013srw}, it is
interesting to derive approximate expressions for it. Several of us
have argued (see e.g.~\cite{Ball:2007ra,Bonvini:2010tp,peraro}) that
accurate approximations to partonic cross sections may be obtained
from knowledge of their $N$ space singularity structure, both at
finite perturbative order, and at the resummed level. Because the
 $N\to\infty$ singularity and the rightmost singularity at finite $N$ are known to all
orders in $\as$ respectively from threshold (Sudakov) and high
energy (BFKL) resummation, if this is indeed the case it is possible
to construct reliable approximations even to very high orders in
$\as$. The possibility of constructing approximations based on the
combination of results from large and small $N$ resummation has also
been considered in~\cite{MUVttbar,Kawamura:2012cr}.

In this paper, we will pursue this idea in the context of Higgs
production in gluon fusion: we will determine the dominant small $N$
and large $N$ singularities up to N$^3$LO from resummation arguments,
and, after testing our methodology against known results up to
NNLO, we will use them to construct a N$^3$LO approximation.

\section{The partonic cross section and its singularities}
\label{sec:joint}

The factorized Higgs production cross section is
\beq\label{eq:xs}
\sigma(\tau,\mh^2) = \tau
\sum_{ij}\int_\tau^1 \frac{dz}{z}\,\Lum_{ij}\(\frac{\tau}{z},\muf^2\)
\frac1z \hat\sigma_{ij}\(z, \mh^2, \as(\mur^2),\frac{\mh^2}{\muf^2},\frac{\mh^2}{\mur^2}\),
\qquad
\tau=\frac{\mh^2}{s},
\eeq
where $\Lum_{ij}(z,\mu^2)$ are the parton luminosities
\beq
\Lum_{ij}(z,\mu^2) = \int_z^1 \frac{dx}x\, f_i\(\frac zx,\mu^2\) f_j(x,\mu^2).
\eeq
We introduce 
coefficient functions $C_{ij}$, defined as
\beq\label{eq:partonic_xs}
\hat\sigma_{ij}\(z,\mh^2, \as(\mur^2),\frac{\mh^2}{\muf^2},\frac{\mh^2}{\mur^2}\)
= z\,\sigma_0\(\mh^2,\as(\mur^2)\) \,C_{ij}\(z,\as(\mur^2),\frac{\mh^2}{\muf^2},\frac{\mh^2}{\mur^2}\),
\eeq
where $\sigma_0$ is the leading order (LO) partonic cross section, so that
the coefficient function is normalized to $\delta(1-z)$ at leading
order:
\beq\label{cfasexp}
C_{ij}(z,\as) = \delta(1-z)\delta_{ig}\delta_{jg} + \as C_{ij}^{(1)}(z) + \as^2 C_{ij}^{(2)}(z) + \as^3 C_{ij}^{(3)}(z) + \Ord(\as^4),
\eeq
and for simplicity, we have suppressed the dependence on
renormalization and factorization scales $\muf,\mur$. In the sequel,
we will concentrate on the gluon fusion subprocess,
while the contribution from other subprocesses will be only briefly
discussed in Section~\ref{sec:apphlhc}, so in most of the discussion
below we will drop the parton indices $ij$, and assume that both the
coefficient function and luminosity refer to the gluon  channel.

Because the cross section Eq.~(\ref{eq:xs}) is a convolution, its
Mellin transform 
\beq\label{sigmeldef}
\sigma(N,\mh^2)\equiv\int_0^1 d\tau \, \tau^{N-2} \sigma(\tau,\mh^2) 
\eeq
factorizes  in terms of the Mellin space luminosity and coefficient
function, respectively defined as
\begin{align} \label{CN}
 \Lum(N) &\equiv \int_0^1 dz \, z^{N-1} \Lum(z) \nonumber \\
 C(N,\as) &\equiv \int_0^1 dz \, z^{N-1} C(z,\as),
\end{align}
according to
\beq\label{sigmelfact}
\sigma(N,\mh^2)=\sigma_0\(\mh^2,\as\) \Lum(N) C(N,\as).
\eeq

While in momentum space the coefficient functions are distributions, if
the Mellin transform integral has a finite convergence abscissa, the 
$N$ space coefficient function is an analytic function of the complex variable
$N$, given by the integral representation Eq.~(\ref{CN}) to the right
of the convergence abscissa, and by analytic continuation elsewhere. 
Therefore, it is fully determined by knowledge of its
singularities. 

The singularity structure of the perturbative expansion
of $C(N,\as)$ is relatively simple. At any perturbative order, the
rightmost singularity is a multiple pole
located at $N=1$~\cite{bfkl}, with further
multiple poles along the real axis 
at $N=0,-1,-2,\ldots$, with
residues of order one (this is also what is found in all known
fixed order calculations);  $\Re N =1$ is the
convergence abscissa of the Mellin transform, and
as $N\to\infty$, $C(N,\as)$ grows as a power
of $\ln N$. While knowledge of the residues of all poles is required
in order to fully determine 
the function $C(N,\as)$, its behavior in the physical region $1\le
\Re N< \infty$ is mostly controlled by the residues of the leading
(rightmost) pole at $N=1$, together with that of the singularity at infinity. 
Both are known from resummation: Sudakov (soft gluon) resummation  
determines to all orders in the
strong coupling the coefficients of
the $\ln^m N$   
terms which control the behavior as $N\to\infty$, while BFKL (high energy)
resummation determines 
the residues of the leading $\frac1{(N-1)^{n}}$
multiple poles. 

This suggests that an approximation of the coefficient function
Eq.~\eqref{CN} may be constructed by simply combining the large $N$
(soft) and small $N$ (high energy) terms,
\beq \label{eq:Capprox}
C_\text{approx}(N,\as)= \Csoft(N,\as)+\Chard(N,\as),
\eeq
where $\Csoft$ contains terms predicted by Sudakov resummation and
$\Chard$ terms predicted by BFKL resummation. It is clear, however,
that this is only correct if the small $N$ singularities, controlled
by $\Chard$, are unaffected by $\Csoft$, while the large $N$ logarithms, 
controlled
by $\Csoft$, are unaffected by $\Chard$. This is clearly nontrivial:
for example, a term proportional to $\ln^m N$ has a cut at
$N=0$, while 
at each fixed order the expected behavior of the coefficient function
is a pole, rather that a cut.
So the approximate expressions for $\Csoft$ and $\Chard$ 
should reproduce
this behavior, with no spurious singularities.

We will show in the sequel that an
approximate expression of the form of Eq.~(\ref{eq:Capprox}) is
possible, but both  $\Csoft$ and $\Chard$ will have to be carefully
constructed. Indeed we will now show
that constructing $\Csoft$ in such a way that the small $N$
singularity structure is preserved, the agreement at large $N$ is
considerably improved. This result may seem surprising, but it is in
fact a consequence of analiticity.

\subsection{Large $N$}
\label{sec:largen}

We first discuss the computation of the large $N$ (soft) part of the
coefficient function.  All contributions to $C(N,\as)$ which do not
vanish as $N\to\infty$ may be computed from Sudakov resummation, using
techniques summarized long ago in Ref.~\cite{Catani:1996yz}. The
resummed coefficient function has the form
\beq
\label{genres}
C_{\rm res}(N,\as)=g_0(\as) 
\exp\left[\frac{1}{\as} g_1(\as\ln N)+g_2(\as\ln N)+ \as g_3(\as\ln N)+\dots\],
\eeq
with
\begin{align}
g_0(\as)&=1+\as g_{0,1}+\as^2 g_{0,2}+\Ord(\as^3),\\
g_i(\lambda)&=\sum_{k=k_{0,i}}^\infty g_{i,k}\lambda^k, 
\quad \text{for }i\geq1, \qquad k_{0,1}=2,\quad k_{0,i\geq2}=1.
\label{eq:gi_res}
\end{align}
Inclusion of all $g_i$ with $1\leq i\leq k+1$ and of $g_0$ up to
order $\as^{k-1}$ gives the next$^k$-to-leading log approximation to $\ln
C_{\rm res}(N,\as)$; it determines the coefficient of all
contributions to the coefficient function of the form $\as^n \ln^m N$
with $2(n-k)+1\le m\le 2n$. This can be extended to $2(n-k)\le m\le
2n$ by also including the order $\as^k$ contribution to $g_0$ .  The
functions $g_1$, $g_2$ and $g_3$ are known exactly, while $g_0$ is
known up to $\Ord(\as^2)$. The function $g_4$ is only known in
part~\cite{MVV,MV}, but the missing information (the $4$-loop cusp
anomalous dimension) only enters at $\Ord(\as^4)$.  We can thus
determine all large $N$ non-vanishing contributions to $C(N,\as)$ up
to $\Ord(\as^2)$, and all logarithmically enhanced contributions (but
not the constant) to $\Ord(\as^3)$.

The accuracy of an approximation to the Higgs production cross section
at the LHC based on the dominance of threshold terms can be
studied~\cite{bfrsaddle} by using the saddle point method to
determine which is the region in $N$ space that gives the bulk of the
contribution to the cross section. It turns out that, despite the fact
that Higgs production at the LHC is far from the kinematic threshold,
partly because of the underlying partonic kinematics and partly
because of the shape of the cross section, at the LHC with 8~TeV
center-of-mass energy, logarithmically enhanced terms are still
providing most of the cross section, though the situation gradually changes as
the center-of-mass energy increases.

However, our goal here is to construct an approximation to the
coefficient function which holds for all $N$ in the physical region.
Now, it has been
observed long ago~\cite{Kramer:1996iq} that the the quality of the
soft approximation to the full coefficient function significantly
depends on the choice of subleading terms which are included in the
resummed result: indeed, while resummation uniquely determines the
coefficients of logarithmically enhanced terms, there is a certain
latitude in defining how the soft approximation is constructed, by
making choices which differ by terms which vanish as $N\to\infty$.
 A similar situation has been observed recently in Drell-Yan production
at the LHC~\cite{Bonvini:2010tp}, for which the threshold
approximation is generally expected to be less good than for Higgs
production. By comparing results which differ by terms which vanish as
$N\to\infty$,
we will now show that several preferred choices for such
subleading terms are favored by the requirement that some aspects of 
the known small $N$
singularity structure of the exact result be reproduced.

In order to outline our strategy, let us work with the simplest
example. Let us first suppose that we know the $N$ space resummed
coefficient function and that we want to extract from
Eq.~\eqref{genres} an approximate expression for the $\Ord(\as)$
coefficient $C^{(1)}(z)$, which is given by~\cite{spira2,Bonciani:2007ex}
\begin{align}\label{nlocf}
C^{(1)}(z) &= 4A_g(z)\, \D_1(z) + d\, \delta(1-z) 
       -2A_g(z) \frac{\ln z}{1-z} + \Rgg(z),\\
\D_k(z) &\equiv \plus{\frac{\ln^k(1-z)}{1-z}},\label{eq:Dk}\\
A_g(z) &\equiv \frac{C_A}{\pi} \frac{1-2z+3z^2-2z^3+z^4}{z}.\label{eq:Ag_def}
\end{align}
The constant $d$ and the function $\Rgg(z)$ are known functions of $\mh/\mt$;
in particular $\Rgg(z)$ is an ordinary function,
regular in $z=1$, so its Mellin transform vanishes as $N\to\infty$
and therefore its specific form
is of no relevance for the large $N$ behavior.

Expanding Eq.~\eqref{genres} to $\Ord(\as)$, and keeping NLL terms, we find
\begin{align}
\label{NLL1}
C_{\rm res}(N,\as)&=1+\as C^{(1)}_{\rm res}(N)+\Ord(\as^2),
\\
C^{(1)}_{\rm res}(N)&=g_{1,2}\ln^2N+g_{2,1}\ln N+g_{0,1},
\label{eq:C1res}
\end{align}
with
\beq
g_{1,2} = \frac{2C_A}{\pi},\qquad
g_{2,1} = \frac{4C_A}{\pi} \gammae,
\eeq
where $\gammae$ is the Euler-Mascheroni constant.
The asymptotic behavior of the $\Ord(\as)$ coefficient as $N\to\infty$
is correctly reproduced by this expression, in that
\beq
\lim_{N\to\infty}\[C^{(1)}_{\rm res}(N)-C^{(1)}(N)\]=0,
\label{Ninfty}
\eeq
where $C^{(1)}(N)$ is the Mellin transform of Eq.~\eqref{nlocf};
the constant $g_{0,1}$ is fixed by this condition.

On the other hand, the behavior of Eq.~\eqref{eq:C1res} at small values of
$N$ is incompatible with the known singularity structure. In
particular, there is a logarithmic branch cut starting at $N=0$ which
is definitely unphysical, as the exact coefficient function has poles
and not cuts at small $N$.  This cut is a subleading singularity, given
that the leading singularity is located at $N=1$, but close enough to
the leading one that the behavior of the coefficient function can be
significantly affected.  Even if we plan to eventually improve this
expression by introducing the correct singularity at $N=1$ according to
Eq.~\eqref{eq:Capprox},
the logarithmic singularity will interfere with it and spoil the accuracy
of the approximation.

This problem, however, is an artifact of the large $N$ approximation,
since powers of $\ln N$ are the
large $N$ approximation of powers of the digamma function $\psi_0(N)$
appearing in fixed order computations. Indeed,
the inverse
Mellin transform of Eq.~\eqref{eq:C1res} (using Eq.~(\ref{eq:Mellin_log2}) of
Appendix~\ref{largenapp}) is seen to be
\begin{align}
C^{(1)}_{\rm res}(z,\as)
&= g_{0,1}\delta(1-z)+2g_{1,2} \Dm_1(z) + \(2\gammae g_{1,2}-g_{2,1}\)\Dm_0(z),
\nonumber\\
&= g_{0,1}\delta(1-z)+\frac{4C_A}\pi \Dm_1(z),
\label{NLLz}
\end{align}
where
\beq\label{eq:DkMP_def}
\Dm_k(z) \equiv \plus{\frac{\ln^k\ln\frac1z}{\ln\frac1z}},
\eeq
which is seen to differ from the soft contribution Eq.~(\ref{eq:Dk}) 
to the exact result Eq.~(\ref{nlocf}).

This can be understood noting that
singular terms as $z\to 1$ arise from integration of the
real emission diagrams over the transverse momentum of  
the gluon, which has the form
\beq\label{eq:soft_emission}
p_{gg}(z)\int_{\Lambda}^{\frac{M(1-z)}{\sqrt{z}}}\frac{dk_T}{k_T}
=\frac{A_g(z)}{1-z}
\(\ln\frac{1-z}{\sqrt{z}}+\ln\frac{M}{\Lambda}\),
\eeq
where $\Lambda$ is a collinear cut-off and $p_{gg}(z)$
is the LO gluon-gluon Altarelli-Parisi
splitting function for $z<1$,
\beq\label{eq:pgg}
p_{gg}(z) = \frac{A_g(z)}{1-z},
\eeq
with $A_g(z)$ given by Eq.~\eqref{eq:Ag_def}.

Indeed, Eq.~\eqref{eq:soft_emission} shows that
logarithmically enhanced soft terms,
rather than being proportional to $\frac{\ln\ln\frac1z}{\ln\frac1z}$, 
are of the form
\begin{equation}\label{sqrtexp}
\frac{1}{1-z}\ln\frac{1-z}{\sqrt{z}}=\frac{1}{1-z}
\Big[\ln(1-z) + \Ord(1-z)\Big],
\end{equation}
and  they appear with a coefficient proportional to the
Altarelli-Parisi splitting function. Explicitly, the latter 
in the $z\to1$ limit may be expanded as
\beq\label{dglapexp}
A_g(z) = \frac{C_A}{\pi} \Big[ 1- (1-z) +2(1-z)^2+ \Ord\[(1-z)^3\] \Big].
\eeq
Logarithmically enhanced contributions to the coefficient function
are generated by the first terms in both expansions 
Eqs.~\eqref{sqrtexp} and \eqref{dglapexp},
namely $\ln(1-z)$ and $A_g(1)$ respectively.

We will now argue that an optimal choice of the soft approximation,
differing from Eq.~\eqref{NLLz} by subleading terms,
is obtained by writing the large soft logs as powers of 
$\ln\frac{1-z}{\sqrt{z}}$,
so in particular retaining the $\sqrt{z}$ in the denominator despite
the fact that it is subleading, and furthermore, by retaining at least
the first correction on the right-hand side of Eq.~(\ref{dglapexp}), also
subleading.
Therefore in this case our suggestion consists in the simple replacement
\beq\label{dk1}
A_g(1)\,\Dm_1(z)\to
A_{g,m}(z)\,\Dh_1(z)
\eeq
in Eq.~\eqref{NLLz}, where $A_{g,m}(z)$ is a finite $m$-th order 
expansion of $A_g(z)$ about $z=1$, Eq.~\eqref{dglapexp}, and
\beq\label{eq:Dh1}
\Dh_1(z) \equiv \plus{\frac{\ln(1-z)}{1-z}}-\frac{\ln\sqrt{z}}{1-z}.
\eeq
Note that we have chosen to apply the plus prescription only to the
first term, singular in $z=1$, which is the natural choice in
fixed order calculations.  In this way, $\Dh_1(N)$ differs from
$\D_1(N)$ only by terms vanishing at large $N$.  Since $\Dm_1(N)$ and
$\Dh_1(N)$ differ at large $N$ by a constant, the coefficient
$g_{0,1}$ must be modified accordingly, in order that the requirement
Eq.~\eqref{Ninfty} be satisfied.  These technical details are
discussed in Appendix~\ref{largenapp}.

Our conclusion Eq.~\eqref{dk1} relies on the following arguments:
\begin{itemize}
\item The replacement of $\Dm_1(z)$, whose Mellin transform is
  \beq
  \Dm_1(N) = \frac12 \[ \ln^2N + 2\gammae\ln N \],
  \eeq
   with $\Dh_1(z)$, whose Mellin transform is
  \beq\label{eq:Dh1N}
  \Dh_1(N) = \frac12 \[\psi_0^2(N)+ 2\gammae\psi_0(N) + \zeta_2 + \gammae^2 \]
  \eeq
  removes the logarithmic branch cut of $\Dm_1(N)$,
  which is incompatible with the known analytic
  structure of the coefficient function. The only
  singularities are now isolated poles, as in the exact
  expression.

\item The same features are shared by the Mellin transform of $\D_1(z)$,
  that is
  \beq\label{eq:D1N}
  \D_1(N) = \frac12 \[ \psi_0^2(N) - \psi_1(N) + 2\gammae\psi_0(N) 
  + \zeta_2 + \gammae^2 \].
  \eeq
  However, the presence of $\psi_1(N)$ exactly cancels the double
  poles of $\psi_0^2(N)$ in $N=0,-1,-2,\ldots$, which are
  there in the exact result.  Therefore, the choice of $\Dh_1(N)$ is
  preferred over $\D_1(N)$.

\item In the replacement Eq.~\eqref{dk1} the factor
  $A_g(z)$ is expanded up to a finite order $m>0$ about $z=1$. 
  This is because the 
  inclusion of the full $A_g(z)$ would introduce a spurious singularity
  in $N=1$. 
  Indeed, the Mellin transform of
  $A_g(z)\,\Dh_1(z)$ is given by
  \begin{multline}
  \int_0^1dz\, z^{N-1}A_g(z)\,\Dh_1(z) \\
  = \frac{C_A}{\pi} \[\Dh_1(N-1)-2\Dh_1(N)+3\Dh_1(N+1)-2\Dh_1(N+2)+\Dh_1(N+3)\].
  \end{multline}
  The first term, due to $1/z$ in $A_g(z)$, has a double and a simple
  pole in $N=1$, while the exact singularity is a simple pole, with a
  $(\mh/\mt)$-dependent coefficient controlled by small $z$
  resummation.
  The expansion of $A_g(z)$ in powers of $1-z$ to any finite order
  is not singular in $z=0$, and therefore does not affect the
  singularity structure around $N=1$.

\end{itemize}

We turn now to the general case. Each of the above arguments can
be generalized to all orders, where $N$ space resummed results contain
powers of $\ln^kN$, whose inverse Mellin transform is a linear combination of 
distributions $\Dm_j(z)$ Eq.~(\ref{eq:DkMP_def}) with $j\leq k-1$.
The fact that the NLO result in $z$ space depends on powers of
$\ln\frac{1-z}{\sqrt{z}}$ rather than $\ln\ln\frac{1}{z}$ is of
kinematical origin and ultimately comes from the upper bound for the
transverse momentum of emitted gluons, Eq.~\eqref{eq:soft_emission},
and therefore it persists to all orders.
It follows that the exact result to all orders is expressed in terms
of distributions $\Dh_k(z)$, defined in Eq.~\eqref{dhatdef}
of Appendix~\ref{largenapp} in analogy with Eq.~\eqref{eq:Dh1}.
The Mellin transform of such distributions, $\Dh_k(N)$,
first, has poles rather than cuts
as small $N$ singularities, and also, in comparison to the
distributions $\D_k(N)$, lacks contributions proportional to powers of
$\psi_k(N)$ with $k$ odd, which would change the pole structure
(see Appendix~\ref{largenapp}).

It has been shown in Refs.~\cite{Kramer:1996iq,Contopanagos:1996nh} 
that the factor $A_g(z)$,
Eq.~\eqref{eq:Ag_def}, is present to all orders, because the full
leading order anomalous dimension exponentiates. However, terms beyond
the first in its expansion Eq.~(\ref{dglapexp}) generate contributions 
 $\as^n(1-z)^j\ln^{2n-1}(1-z)$ with $j\ge 0$ to the coefficient
functions,
which are generally of the same order as other terms which we do not
control. However, it can be shown~\cite{Catani:2001ic} that 
the inclusion of the
$\Ord\[(1-z)^1\]$ term in the expansion Eq.~\eqref{dglapexp} correctly
predicts, after exponentiation, the subdominant contributions of the
form $\as^n\ln^{2n-1}(1-z)$ (i.e., in $N$ space, terms behaving as
$\as^n N^{-1}\ln^{2n-1}N$ at large $N$) to all orders, so the
inclusion of this term rests on firm ground.

Including the $\Ord\[(1-z)^1\]$  from Eq.~\eqref{dglapexp} we get 
\beq\label{eq:Ag_order1}
A_{g,1}(z) = \frac{C_A}{\pi}\[1-(1-z)\] = z\, A_g(1), 
\eeq
which is easily implemented to all orders by the
replacement
\beq
\label{eq:Dlog->Dhat_N} 
\Dm_k(z)\to z\,\Dh_k(z);\qquad
\Dm_k(N) \to \Dh_k(N+1).  
\eeq

Including also the next  order gives
\beq
A_{g,2}(z) = \frac{C_A}{\pi}\[1-(1-z) + 2(1-z)^2\]
 = \[2-3z+2z^2\] A_g(1),
\eeq
which amounts to replacing
\beq\label{eq:Dlog->Dhat_N2}
\Dm_k(N) \to 2\Dh_k(N) -3\Dh_k(N+1) +2\Dh_k(N+2),
\eeq
in the $N$ space expressions.
The third order term of the expansion of $A_g(z)$ is accidentally zero,
so $A_{g,2}(z)=A_{g,3}(z)$. We have checked that
the inclusion of terms of order $(1-z)^4$ and higher
in the expansion of $A_g(z)$ does not affect our results significantly.
We 
will consider both the expansions to first and second order, and use
their difference as a means to estimate the uncertainty on the
result. Specifically, we will take the mid-point between them as our
best prediction, with the first- and second-order expansion result
giving the edges of the uncertainty band.

In summary, our soft approximation (to be combined with small $N$
terms determined in the next Section) is constructed in the following
way.  The resummed expression Eq.~\eqref{genres} can be rewritten
\beq\label{eq:CresDlog} 
C_{\rm res}(N,\as) = g_0(\as) \,\exp
\sum_{n=1}^\infty \as^n \sum_{k=0}^{n} b_{n,k}\,\Dm_k(N), 
\eeq 
where
the coefficients $b_{n,k}$ are obtained from the functions $g_i$,
Eq.~\eqref{eq:gi_res}, and have been determined up to $n=3$~\cite{MV}.
The function $g_0(\as)$ is known only up to $\Ord(\as^2)$; the 
uncertainty associated to $g_{0,3}$ will be discussed in Sect.~\ref{sec:partonic}.

The replacements Eq.~\eqref{eq:Dlog->Dhat_N} or
\eqref{eq:Dlog->Dhat_N2} are then applied to Eq.~\eqref{eq:CresDlog}.
We obtain, respectively,
\begin{subequations}\label{eq:Csoft}
\begin{align}
\label{eq:Csoft1}
\Csoftm{1}(N,\as) 
&= \bar g_0(\as) \,
\exp \sum_{n=1}^\infty \as^n \sum_{k=0}^{n} b_{n,k}\,\Dh_k(N+1), \\
\label{eq:Csoft2}
\Csoftm{2}(N,\as)
&= \bar g_0(\as) \,\exp \sum_{n=1}^\infty \as^n \sum_{k=0}^{n}  b_{n,k}\,
\[2\Dh_k(N) -3\Dh_k(N+1) +2\Dh_k(N+2)\],
\end{align}
\end{subequations}
where we have defined
\begin{align}
\label{eq:g0bardef}
\bar g_0(\as) &=g_0(\as) \exp\[-\sum_{n=1}^\infty \as^n \sum_{k=0}^{n}  b_{n,k}d_k\]
\\
\label{eq:dk_def}
d_k&=\lim_{N\to\infty}\[\Dh_k(N)-\Dm_k(N)\]
\end{align}
so that the condition Eq.~\eqref{Ninfty} is satisfied to all orders
after the replacement. Explicit expressions for the coefficients
$b_{n,k}$ and $d_k$ are given in Appendix~\ref{largenapp}.

Equations~\eqref{eq:Csoft} can be cast in the form
\beq
\Csoft(N,\as) = \bar g_0(\as) \,\exp \sum_{n=1}^\infty \as^n\, S_n(N),
\eeq
which is now expanded in powers of $\as$:
\beq\label{Csoft}
\Csoft(N,\as) = 1 + \as \Csoft^{(1)}(N) + \as^2 \Csoft^{(2)}(N)
+ \as^3 \Csoft^{(3)}(N) +\Ord(\as^4).
\eeq
We obtain
\begin{subequations}\label{Csoftexp}
\begin{align}
\Csoft^{(1)}(N) &= S_1(N) +\bar g_{0,1}\\
\Csoft^{(2)}(N) &= \frac12 S_1^2(N) + S_2(N) + \bar g_{0,1}S_1(N) + \bar g_{0,2}\\
\Csoft^{(3)}(N) &= \frac16 S_1^3(N) + S_1(N)S_2(N) + S_3(N) + \bar g_{0,1}\(\frac12 S_1^2(N)+S_2(N)\)
  + \bar g_{0,2} S_1(N) + \bar g_{0,3}.
\end{align}
\end{subequations}

As a test of our procedure we now compare the first two orders of our
soft approximations Eq.~\eqref{eq:Csoft} to the full result.  Note that
in the sequel when comparing to known results, and also when
constructing our $\Ord(\as^3)$ approximation, we will always be
retaining the exact $\mt$ dependence.

As terms of comparison, at NLO we use the finite-$\mt$ result of
Ref.~\cite{spira2} (using the numerical implementation of
Ref.~\cite{Bonciani:2007ex}), while at NNLO we use the approximate
finite-$\mt$ result obtained by matching the double expansion in
powers of $1-z$ and $\mh/\mt$ of
Refs.~\cite{harlander1,harlander2} to the known small $z$ terms
computed in Ref.~\cite{Higgsfinite} according to
Ref.~\cite{harlander3} (see Refs.~\cite{pak1,pak2} for further
approximate finite-$\mt$ results). Note that the soft limit only
depends on $\mt$ through the function $g_0(\as)$ of Eq.~\eqref{genres}.

\begin{figure}[t]
\centering
\includegraphics[width=0.495\textwidth,page=1]
{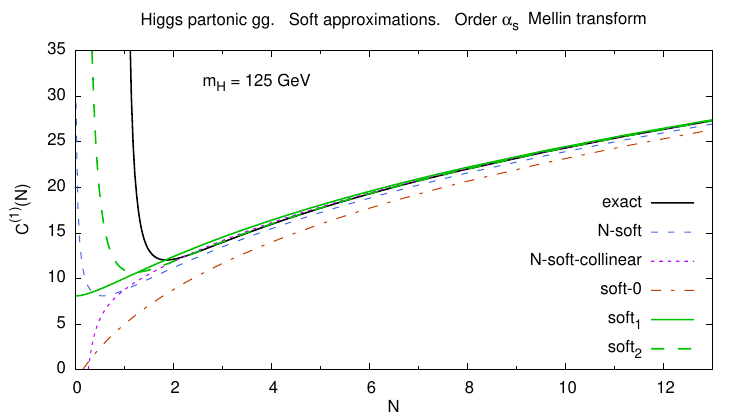}
\includegraphics[width=0.495\textwidth,page=1]
{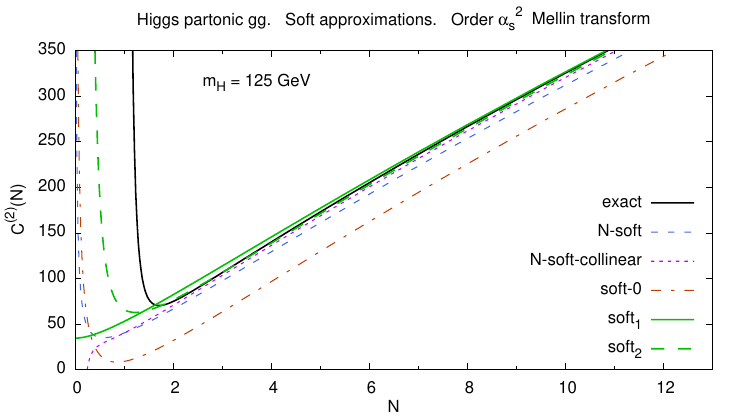}
\caption{The partonic coefficient function Eq.~\eqref{Csoft} in $N$
  space for $\mh=125$~GeV at NLO (left) and NNLO (right) and its soft
  approximation in various forms: our preferred choices
  Eqs.~\eqref{eq:Csoft}, denoted as soft$_1$ and soft$_2$,  the
  simpler approximations based on $\Dm(N)$ as in Eq.~\eqref{eq:CresDlog},
  denoted as $N$-soft, its collinear-improved version from
  Ref.~\cite{Catani:2003zt}, 
and the approximation based on  $\D(N)$, as in
  Eq.~\eqref{dsimple}, denoted as soft-$0$.}
\label{fig:soft_comparison}
\end{figure}

Results are shown, as functions of $N$ along the real $N$ axis, in
Fig.~\ref{fig:soft_comparison}. We find the comparison in $N$ space to
be most instructive, because the coefficient function is then an
ordinary function, rather than a distribution as in $z$
space. Furthermore, the saddle point which dominates the Mellin
inversion is on the real axis~\cite{bfrsaddle}. All this said, it
should be kept in mind that the physical cross section is obtained by
Mellin inversion of the product of the $N$ space coefficient function
and luminosity: therefore, agreement on the real axis is certainly
necessary, but in general not sufficient for agreement of the physical
results. In particular, spurious singularities (and in particular
spurious cuts) may substantially modify the behavior of the
coefficient function in the complex plane.

In order to understand the role of various subleading terms, we also
show in Fig.~\ref{fig:soft_comparison} the results obtained 
expanding the resummed expression, Eq.~\eqref{eq:CresDlog}, which is
built up from the distributions $\Dm_k(z)$, Eq.~\eqref{eq:DkMP_def},
and thus it has spurious cuts starting at $N=0$ (labeled
 $N$-soft),
and the one obtained expanding the resummed expression Eq.~\eqref{eq:CresDlog}
in powers of $\as$ and then replacing
\beq\label{dsimple}
\Dm_k(N) \to \D_k(N)
\eeq
(and adjusting the constant term),
where $\D_k(N)$ are the Mellin transforms of the $\D_k(z)$ distributions,
Eq.~\eqref{eq:Dk}, so it does not include the contributions coming
from the $1/\sqrt{z}$ in the phase space integration (labeled
 soft-$0$). We finally show the results found expanding out the
collinear-improved resummed results of Ref.~\cite{Catani:2003zt}, 
(labeled $N$-soft-collinear): these differ from the $N$-soft curves
by the addition to $g_0$ Eq.~(\ref{genres}) of an $\Ord(1/N)$
contribution of collinear origin. This is akin to the collinear improvement
which is effected in our result by the shift
Eq.~(\ref{eq:Dlog->Dhat_N}): indeed, the  subdominant $\as^n
N^{-1}\ln^{2n-1}N$ contributions (which as mentioned above are
universal) generated by this collinear improvement coincide with those
which we also include through our shift.

While our preferred options clearly provide the best approximation to
the exact result in the soft region $N\gtrsim2$, it is interesting to
observe that the $N$-soft form, based on
Eq.~\eqref{eq:CresDlog}, despite having the wrong singularity
structure (cuts rather than poles at small $N$), still provides a
reasonable approximation, though not quite as good as our preferred
ones.  This can be understood noting that~\cite{Bonvini:2010tp}
\beq
\frac{\ln^k\ln\frac{1}{z}}{\ln\frac{1}{z}}
=\frac{\sqrt{z}}{1-z}\ln^k\frac{1-z}{\sqrt{z}} 
\times\Big[1+\Ord\[(1-z)^2\]\Big],
\eeq
which means that
$\Dm_k(z) =\sqrt{z}\, \Dh_k(z)+\text{delta terms}+\Ord\[(1-z)\ln^k(1-z)\]$,
i.e., this choice for the logarithms is very similar to our $\Csoftm{1}$
approximation Eq.~\eqref{eq:Csoft1}, up to a shift in $N$ by
$1/2$. The $N$-soft-collinear result is quite close to the $N$-soft,
to which it approaches at small $N$,
but somewhat closer to our own, especially at large $N$ (by construction the
$N$-soft and the
$N$-soft-collinear results coincide when  $N=1$).

\subsection{Small $N$}
\label{sec:smalln}

The leading small $N$ singularities for the Higgs inclusive cross
section have been determined to all orders in $\as$ in
Ref.~\cite{hautmannHiggs} in the $\mt\to\infty$ limit, and in
Ref.~\cite{Higgsfinite} for finite $\mt$. These results have been
obtained by means of the so-called high energy  or $k_t$ factorization
technique of Ref.~\cite{cch}, which has been subsequently used to
compute high energy cross section for an increasing number of
processes~\cite{ch,ball_ellis,ball_marz,diana,caola_marz} and more
recently extended to rapidity distributions (and also used to
determine all order results for Higgs production) in Ref.~\cite{cfm}.

In this formalism, small $N$ singularities are obtained to all orders
by computing the leading order partonic cross section for the relevant
process, but with off-shell incoming gluons. They are extracted from
the off-shell coefficient function $C_\text{off-shell}$, defined through
\beq
\hat{\sigma}_{g^* g^*\to H} = z\, \sigma_0 \,C_\text{off-shell}(z, \xi_1,\xi_2) ,
\eeq
where $z$ is the scaling variable defined in the previous section,
while $\xi_i=|{\kt}_i|^2/\mh^2$ in terms of the off-shellness
$k_i^2=-|{\kt}_i|^2$ of the $i$-th incoming gluon, and the angle
between the incoming transverse momenta is integrated over.
To do this, one defines the
impact factor $h(N,M_1,M_2)$ according to
\begin{align} \label{impact}
h(N,M_1,M_2) &= M_1\, M_2\, R(M_1)\, R(M_2) \(\frac{\mh^2}{\muf^2}\)^{M_1+M_2}
 \nonumber \\ & \qquad \times
\int_0^1 \frac{dz}z\,z^{N} \int_0^\infty \frac{d\xi_1}{\xi_1}\,\xi_1^{M_1} \int_0^\infty \frac{d\xi_2}{\xi_2}\,\xi_2^{M_1}
\, C_\text{off-shell}(z, \xi_1,\xi_2) \nonumber \\
&= \sum_{i_1,i_2=0}^\infty c_{i_1,i_2}(\mt,\mh,\muf)\, M_1^{i_1}
M_2^{i_2}+ \Ord\left(N-1 \right),
\end{align}
where the pre-factor $R$ accounts for
factorization scheme dependence~\cite{cch2}, and in $\MSbar$
is given by
\beq \label{cataniR}
R(M) = 1+\frac83\zeta_3 M^3 + \Ord(M^4).
\eeq
The determination of the coefficients $c_{i_1,i_2}(\mt,\mh,\muf)$ has
been reduced to quadratures to all orders in Ref.~\cite{Higgsfinite};
they have 
been numerically determined up to and including second order in $\as$ in~\cite{Higgsfinite} and
up to and including fourth order in~\cite{marzaniPhD}.

The leading singularities of the partonic coefficient function
are obtained by identifying the Mellin variables $M_i$ with the
anomalous dimension $\gamma^+_s$. This, in turn, is the eigenvalue of
the singlet anomalous dimension matrix which contains, to all orders in $\as$,
the contributions with the highest powers of the rightmost $N$ space
singularities. Indeed, as well known, only one of the two
eigenvalues (which henceforth we will refer to  as the
``large'' eigenvalue) has singularities at $N=1$,\footnote{Note that in the small $N$ literature, and
 specifically in Refs.~\cite{Higgsfinite,marzaniPhD} the
 variable $N$ is usually shifted by one unit, so that the singularities
 of $\gamma^+$ 
 are located at $N=0$, by taking $x^N$ instead of $x^{N-1}$ as a
 kernel of the Mellin transform Eq.~(\ref{CN}). Throughout this paper
 we adopt instead the more common convention of  Eq.~(\ref{CN}).} 
while the other has
singularities at $N=0$.

In other words, the leading singularities are found letting $M_i=\gamma^+_s$, with
\beq \label{gammas}
\gamma^+_s =\sum_{n=1}^\infty e_{n-1,\,-n} \left(\frac{\as(\mh^2)}{N-1} \right)^n
\eeq
where the coefficients $e_{n-1,\,-n}$ are determined~\cite{Jaroszewicz:1982gr} using
duality~\cite{Ball:1999sh} from the leading order BFKL kernel. The
first 35 coefficients $e_{n,\,-n}$ are tabulated in Ref.~\cite{Ball:1995vc};
the first few have accidental zeros, and are given by $e_{0,-1}= C_A/\pi$,
$e_{1,-2}=e_{2,-3}=0$, $e_{3,-4}=2\zeta_3(C_A/\pi)^4$, $e_{4,-5}=0$.
It follows that to $k$-th order in $\as$ the coefficient function has
a $k$-th order pole in $N=1$. Note that in the heavy top limit the
small $N$ singularity structure is different, in that at each extra
order in $\as$ the order of the pole increases by two
units~\cite{hautmannHiggs}. However, these double poles are unphysical, and follow
from a breakdown of the large $\mt$ approximation at high energy: we
will thus not
discuss them further.

It has been shown in Refs.~\cite{Ball:2007ra, abfquarks} that the nature
of the small $N$ singularity of coefficient functions at the resummed
level is entirely determined by the singularity of the resummed
anomalous dimension $\gamma^+$. However, reproducing the correct
all order small $N$ singularity of the anomalous dimension (which is a
simple pole to the right of $N=1$, but close to it)
requires~\cite{abffinal} the all order inclusion of two classes of
subleading terms on top of the leading (or next-to-leading)
singularities Eq.~(\ref{gammas}): namely, running coupling
corrections, without which the small $N$ leading singularity would be
a square-root cut instead of a simple pole~\cite{Altarelli:2001ji},
and anticollinear terms~\cite{Salam:1998tj} without which the
perturbative expansion of both the position and residue of the above
simple pole would not be stable (similar conclusions can also be
arrived at from a study~\cite{Ciafaloni:2002xf,ccss} of the
BFKL~\cite{bfkl} equation).  The inclusion of a further series of
all order running coupling corrections in the coefficient function is
further required~\cite{Ball:2007ra, abfquarks} in order for this not to develop
extra spurious singularities.  When expanded out in perturbation
theory, these running coupling corrections correspond to series of
contributions of increasingly low logarithmic order (i.e.\
increasingly subleading): we will retain both up to the
NLL order, i.e.\ keeping not only the leading
singular contribution to each order in $\as$, but also the first
subleading correction, i.e.\ to order $\as^k$ both the contributions
with a $k$-th and a $(k-1)$-th order pole in $N=1$.

For anomalous dimensions this is simply done by including the full
next-to-leading singular contribution to them.
For coefficient functions, these running coupling corrections 
are found by letting, in
Eq.~(\ref{impact}), 
$M_i^k=\[{\gamma^+_{\rm res}}^k\]$, with $\[{\gamma_{\rm res}^+}^{k}\]$ given recursively by~\cite{Ball:2007ra, abfquarks}
\beq
\[{\gamma_{\rm res}^+}^{k+1}\] = \gamma_{\rm res}^+\(1+k\frac{\dot
 \gamma_{\rm res}^+}{{\gamma_{\rm res}^+}^2}\)\[{\gamma_{\rm res}^+}^k\], \qquad \[{\gamma_{\rm res}^+}\] =
{\gamma_{\rm res}^+}, \eeq
where 
\begin{equation}\label{rcad}
{\dot{\gamma}^+_{\rm res}} =
-\beta_0\as^2\frac{\partial}{\partial\as}{\gamma_{\rm res}^+},
\end{equation}
and with $\gamma_{\rm res}^+$ we have denoted a form of the large
eigenvalue which includes at least the leading singularities
Eq.~(\ref{gammas}), but may include other subleading contributions.

We can now compute the small $N$ approximation to the coefficient
function. We expand the anomalous dimension
to  fixed perturbative order
\beq \label{gpexp}
\gamma^+= \as \gamma^{(0)}+\as^2 \gamma^{(1)}+ \as^3 \gamma^{(2)}+\Ord(\as^4).
\eeq
The leading and next-to-leading singularities of the anomalous dimension $\gamma^+$ are given by
\begin{subequations}
\label{eq:explicitgamma}
\begin{align}
\gamma^{(0)}&= \frac{e_{0,-1}}{N-1}+e_{0,0}+\Ord\left(N-1\right)\\
\gamma^{(1)}&= \frac{e_{1,-2}}{(N-1)^2}+\frac{e_{1,-1}}{N-1}+\Ord\left(1\right)\\
\gamma^{(2)}&= \frac{e_{2,-3}}{(N-1)^3}+\frac{e_{2,-2}}{(N-1)^2}+\Ord\left((N-1)^{-1}\right),
\end{align}
\end{subequations}
where the coefficient of the leading poles can be read off
Eq.~\eqref{gammas}: $e_{0,-1}=\frac{C_A}{\pi}$ and
$e_{1,-2}=e_{2,-3}=0$. The other coefficients are:
\begin{subequations}
\begin{align}
e_{0, 0}&= \frac{-11C_A+2n_f(2C_F/C_A-1)}{12\pi},\\
e_{1,-1}&= \left(\frac{13 C_F}{18 \pi ^2}-\frac{23 C_A}{36 \pi ^2}\right) n_f, \\
e_{2,-2}&= \frac{C_A^3\zeta_3}{2\pi^3} + \frac{11C_A^3\zeta_2}{12\pi^3} - \frac{395 C_A^3}{108 \pi^3}
+\(\frac{C_A^2\zeta_2}{6\pi^3}-\frac{71C_A^2}{108\pi^3}-\frac{C_FC_A\zeta_2}{3\pi^3} +\frac{71C_FC_A}{54\pi^3}\) n_f.
\end{align}
\end{subequations}

The $N\to 1$ result for the partonic coefficient function in the
gluon  channel can be then obtained by substituting
Eq.~\eqref{gpexp} into Eq.~\eqref{impact}:
\begin{align} \label{Ch}
\Cabf(N,\as) &=\sum_{n=1}^\infty \as^n \Cabf^{(n)}(N) \nonumber \\
&= \sum_{i_1,i_2 \ge0} c_{i_1,i_2} \[{\gamma^+}^{i_1}\] \[{\gamma^+}^{i_2}\] -1\nonumber \\
&= \as 2c_{1,0} \gamma^{(0)}\nonumber\\
&\quad+ \as^2\[ \(2 c_{2,0}+c_{1,1}\) {\gamma^{(0)}}^2
- 2 c_{2,0} \beta_0\gamma^{(0)}
+ 2 c_{1,0} \gamma^{(1)} \] \nonumber\\
&\quad+\as^3 \Big[
\(c_{3,0}+c_{2,1}\) 2 {\gamma^{(0)}}^3 - \(3c_{3,0}+c_{2,1}\) 2\beta_0 {\gamma^{(0)}}^2 +4c_{3,0}\beta_0^2\gamma^{(0)}
 \nonumber \\
&\qquad\qquad
+\(2 c_{2,0}+c_{1,1}\)2\gamma^{(0)}\gamma^{(1)} - 4c_{2,0}\beta_0\gamma^{(1)} +2 c_{1,0}\gamma^{(2)} 
\Big]\nonumber\\
&\quad+\Ord\left( \as^4 \right),
\end{align}
where we have omitted the dependence of the coefficients on $\mt$,
$\mh$ and $\muf$ for simplicity. 

\begin{figure}[t]
 \centering
 \includegraphics[width=0.495\textwidth,page=2]{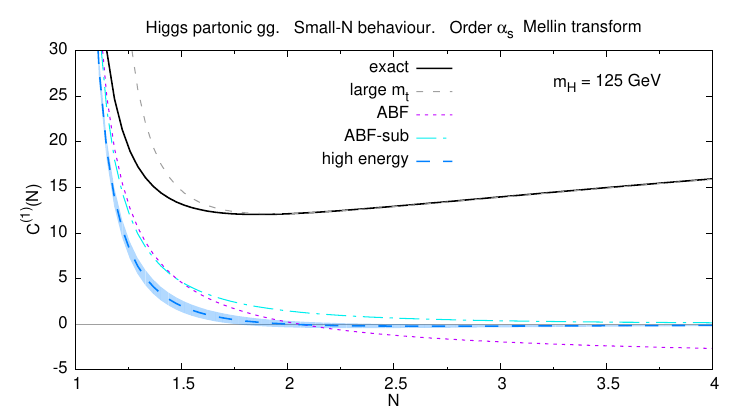}
 \includegraphics[width=0.495\textwidth,page=2]{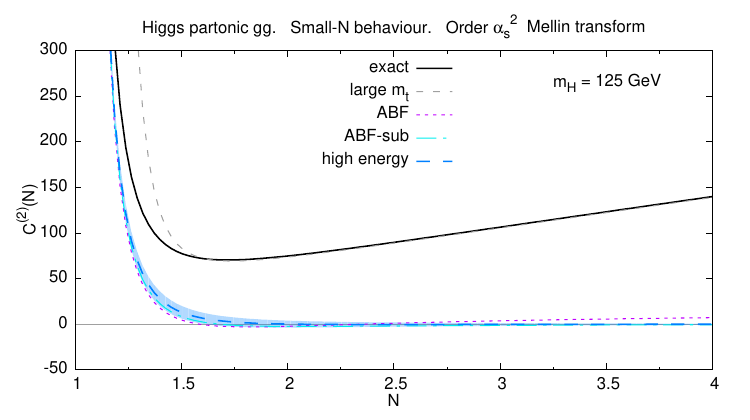}
 \caption{Comparison of the NLO (left) and NNLO (right) exact
   coefficient function both for finite $\mt$ (exact) and in the
   $\mt\to\infty$ limit (large $\mt$), and
 several small $N$ approximations to it: the NLO and NNLO
 contributions to $\Cabf$ Eq.~\eqref{Ch} (ABF), to $\Csub$
 Eq.~\eqref{eq:Charddamp} (ABF-sub) and to
   $\Chard$ Eq.~\eqref{eq:Chardmom} (high energy).}
 \label{fig:smallN_coeff_funct}
\end{figure}


Because we wish to combine the small $N$ behavior which we are
determining here with the large $N$ behavior determined in
Sect.~\ref{sec:largen},  we must make sure that  the small $N$
contribution to the coefficient function vanishes as
$N\to\infty$. However, 
the coefficient function $\Cabf(N)$ Eq.~(\ref{Ch}) manifestly
does not vanish in the large $N$ limit, because of the constant
contribution to $\gamma^{(0)}$ Eq.~(\ref{eq:explicitgamma}) which
propagates into $\Cabf(N)$  to all orders in $\as$.

Therefore, we construct an improved small $N$ approximation to the coefficient
function as a subtracted version of $\Cabf(N)$. The subtracted
coefficient function has the same leading small $N$ singularities as
$\Cabf(N)$, but it vanishes as $N\to\infty$. It is given by
\beq \label{eq:Charddamp}
\Csub^{(n)}(N)= \Cabf^{(n)}(N)-2 \Cabf^{(n)}(N+1)+ \Cabf^{(n)}(N+2).
\eeq
It is apparent that $\Csub^{(n)}(N)$ and $\Cabf^{(n)}$ have
the same leading $N=1$
singularities: the subtraction  only introduces subleading $N=0$ and $N=-1$
singularities. However,  $\lim_{N\to\infty}\Csub^{(n)}(N)=0$. Of
course, many forms of the subtraction are possible: the particular one
given in Eq.~(\ref{eq:Charddamp}) has been chosen as a compromise
between the contrasting goals of not changing the small $N$ behavior
and of damping strongly enough at 
large $N$.  In $z$ space, the subtraction Eq.~(\ref{eq:Charddamp})
corresponds to damping the $z\to1$ behavior of the coefficient
function
through a multiplicative
factor $(1-z)^2$.

In view of combining the small and large $N$ approximations to the
coefficient function, one may ask what is the expected transition
point between the two approximations. In order to answer the question,
a relevant observation is to note that momentum conservation implies
that $\gamma^+(2)=0$ to any order in perturbation theory. This in
particular implies that all $\Cabf^{(n)}(N)$ vanish at
$N=2$. This suggests that $N=2$, which is a fixed point for the
anomalous dimension, marks the transition between the small $N$
approximation (not accurate when $N\gsim 2$) and the large $N$
approximation (not accurate when $N\lsim 2$). In particular, because
the coefficient function Eq.~(\ref{Ch}) is a polynomial in
$\gamma^+(N)$, it vanishes at $N=2$ if the anomalous dimension
does. 

However, the small $N$ approximation 
Eq.~(\ref{gpexp}) to the anomalous dimension
does not respect momentum conservation, because it only includes the
contribution to  $\gamma^+$ from the
leading and next-to-leading singularities in $N=1$
Eq.~\eqref{eq:explicitgamma}, and not the full  
fixed order expression of $\gamma^+$
Eq.~\eqref{eq:explicitgamma}. Momentum conservation can be
enforced~\cite{abfquarks} by
adding to $\Cabf^{(n)}(N)$ a function $f_\text{mom}(N)$.
This function must not introduce spurious singularities
at $N=1$ and it should also be subdominant with respect to the large
$N$ contributions that we control in $\Csoft(N)$. A natural
choice appears to be $f_\text{mom}(N)= c/N$, with $c$ fixed
so that, after subtraction Eq.~\eqref{eq:Charddamp},
our small $N$ coefficient function vanishes in $N=2$. With this
choice, the  small $N$ approximation of the coefficient function becomes
\begin{align} \label{eq:Chardmom}
&\Chard^{(n)}(N)= \Csub^{(n)}(N)- \frac{4 !\, k_{\rm mom}}{N (N+1)(N+2)}.
\end{align}

Note however that the exact coefficient function does not in
general vanish at $N=2$, only the contribution to it driven by hard
radiation from external legs and expressed in terms of the anomalous
dimension does. Thus, for instance, contributions from 
subdominant poles in $N=0, -1, -2, \ldots$ will in general lead to
a non-vanishing contribution to the coefficient function in $N=2$.
Because we do not control such a contribution, we estimate it by
allowing the coefficient function to deviate from zero at $N=2$, by
modifying the value of the constant in the subtraction term of
Eq.~(\ref{eq:Chardmom}). We take this deviation from zero to reach as
its maximum value 5\% of the size of the soft contribution
Eq.~\eqref{Csoft} at $N=2$, $\Csoft(2)$, with
either sign; namely, we choose in Eq.~(\ref{eq:Chardmom})
\beq\label{momconst}
k_{\rm mom}=\Csub(2) \pm 0.05\times \Csoft(2).
\eeq
This means that the small $N$ contribution, rather than
being completely switched off at $N=2$,  is small at that point, and gets
switched off somewhere in its vicinity.

Our final result Eq.~(\ref{eq:Chardmom}) for the small $N$
contribution $\Chard^{(n)}(N)$ to the coefficient function, as well as
several small $N$ approximations are compared to each other and to the
known full result at NLO and NNLO in Fig.~\ref{fig:smallN_coeff_funct}. 
The full result is shown both in the pointlike approximation
(labeled as  large $\mt$),
and for finite $\mt$  (labeled as exact): the different small $N$
behavior of
the pointlike result, due to spurious double poles, is apparent.
The small $N$ approximations Eq.~(\ref{Ch})  (labeled AFB), and its
subtracted version  Eq.~\eqref{eq:Charddamp}  (labeled ABF-sub) are seen
to provide an equally good approximation to the exact result in the
very small $N$ region where the latter is dominated by its small $N$
poles, but only the subtracted version vanishes at large $N$.
The final result Eq.~(\ref{eq:Chardmom})  after enforcing
momentum conservation of the anomalous dimension is finally
shown (labeled high energy), with an uncertainty band obtained
by varying the size of $\Chard^{(n)}(2)$ about zero as discussed
above: it coincides with the small $N$ approximation for $1\leq N\lsim
1.25$, but it is gradually switched off for larger $N$ until vanishing
in the vicinity of $N\sim 2$.

\section{Approximate   cross sections up to N$^3$LO}
\label{sec:partonic}

\subsection{Parton level results}
\label{sec:partonicpl}
We can now construct an approximation to the full coefficient
function. Having constructed a large $N$ approximation $\Csoft(N,\as)$ 
Eq.~(\ref{Csoft}) and a small $N$ approximation $\Chard(N,\as)$ 
Eq.~(\ref{eq:Chardmom}) to the coefficient function, in such a way
that the small $N$ term does not spoil the large $N$ singularities and
conversely, we can combine them using
 Eq.~(\ref{eq:Capprox}), which we then expand out in powers of
 $\alpha_s$ according to Eq.~(\ref{cfasexp}), so that at N$^k$LO we
 have
\beq \label{approxfinal}
C^{(k)}_\text{approx}(N)= \Csoft^{(k)}(N)+\Chard^{(k)}(N).
\eeq

\begin{figure}[t]
\centering
\includegraphics[width=0.495\textwidth]{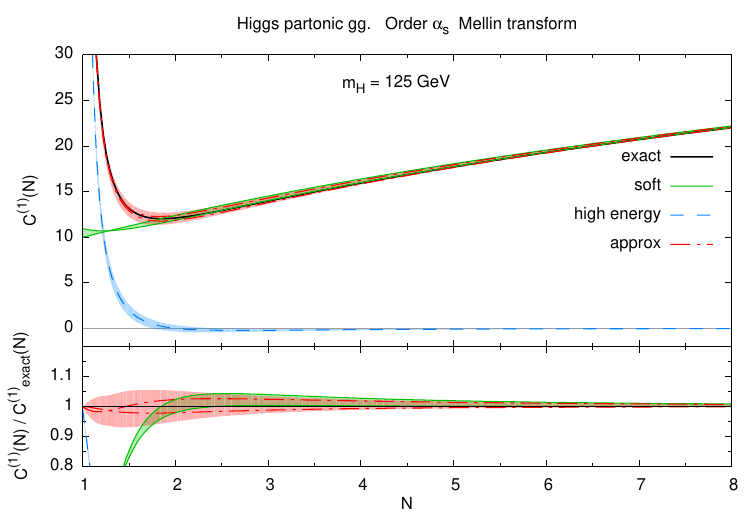}
\includegraphics[width=0.495\textwidth]{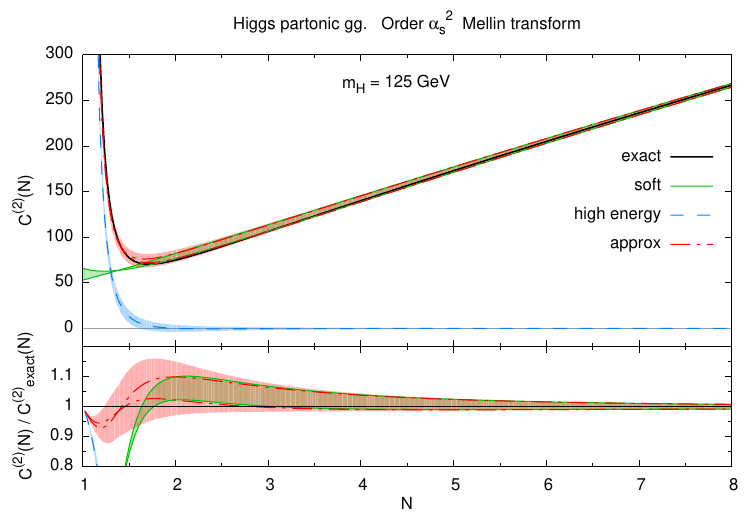}
\caption{Comparison of the NLO (left) and NNLO (right) exact
  coefficient functions to various approximations to it. The
   large $N$ approximation, corresponding to the band
  between the soft$_1$ and soft$_2$ curves in
  Fig.~\ref{fig:soft_comparison} (soft); the small
  $N$ approximation,  corresponding to the high energy curve in
  Fig.~\ref{fig:smallN_coeff_funct} (high energy);
  and the combined small and large $N$
  approximation Eq.~(\ref{approxfinal}) (approx). 
  The bottom plot shows the ratio of the approximate results
  to the exact result. Note that at NNLO the ``exact'' result is in
  fact the approximate construction of Ref.~\cite{harlander1,harlander2}.}
 \label{fig:approx_coeff_funct}
\end{figure}

Before turning to the N$^3$LO, which is our main result, 
 we first compare the NLO and NNLO results found using our procedure
 to the corresponding exact results.
We will use $\mh=125$~GeV and $\mt=172.5$~GeV throughout.
The comparison is shown in Fig.~\ref{fig:approx_coeff_funct}, where
our best approximate result $C^{(k)}_\text{approx}(N)$ Eq.~(\ref{approxfinal})
(labeled as approx) is shown
along with the large $N$ $\Csoft^{(k)}(N)$ (labeled as soft) and small $N$
$\Chard^{(k)}(N)$ (labeled as high-energy) terms which contribute to it.  
As discussed in Sect.~\ref{sec:largen} and Sect.~\ref{sec:smalln}
respectively, the uncertainty on $\Csoft(N,\as)$ is obtained as the
spread between the two different forms Eq.~(\ref{eq:Csoft})  of the
large $N$ approximation (green band), while the uncertainty on $\Chard(N,\as)$ is
obtained by varying the size of $\Chard^{(n)}(2)$ about zero (blue band).
The uncertainty on $C^{(k)}_\text{approx}(N)$ Eq.~(\ref{approxfinal})
is then obtained as the envelope of these uncertainty bands (red band).
In each plot we also show the ratio of the approximate result to the
exact one.

It is apparent that the approximate results reproduce the exact one 
within the uncertainty in the full
region of real $N>1$ at NLO, while at NNLO there is a small disagreement
(of about $5\%$) very close to $N=1$. Note, however,
that in this region what we call ``exact'' result is not necessarily reliable:
indeed, in the absence of a full NNLO result we are taking
as exact the  matching of Ref.~\cite{harlander1,harlander2} 
of a double expansion in
powers of $1-z$ and $\mh/\mt$
 with the exact leading (double) $N=1$ pole
computed in Ref.~\cite{Higgsfinite}. In particular, the contribution
from subleading poles (single pole at $N=1$ and multiple poles for
non-positive integer $N$) in the ``exact'' result are not correctly
reproduced, while in our approximate expression
they are partly estimated by varying the
size of $\Chard^{(n)}(2)$.

\begin{figure}[t]
 \centering
 \includegraphics[width=0.7\textwidth,page=2]{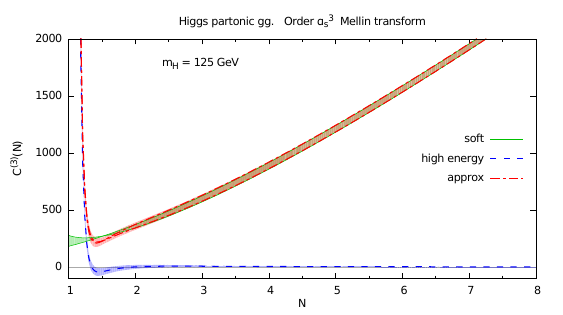}
 \caption{Same as Fig.~\ref{fig:approx_coeff_funct}, but at N$^3$LO order.}
 \label{fig:approx_C3}
\end{figure}

We now consider our new result for the N$^3$LO coefficient function:
$\Csoft^{(3)}(N)$ is given either by Eq.~\eqref{eq:Csoft1} or Eq.~\eqref{eq:Csoft2},
and $\Chard^{(3)}(N)$ is given in Eqs.~\eqref{eq:Charddamp},
\eqref{eq:Chardmom} in terms of $\Cabf^{(3)}(N)$ Eq.~(\ref{Ch}).
The coefficients in the large $N$ contribution 
Eqs.~\eqref{eq:Csoft} are collected
in Appendix~\ref{largenapp}, except the coefficient $\bar g_{0,3}$, which is
unknown: unless stated otherwise, the results are presented with $\bar
g_{0,3}=0$. This is a coefficient  in the expansion of the constant
function $\bar g_0(\as)$, related by  Eq.~(\ref{eq:g0bardef}) to the
function  $g_0(\as)$ which appears in the resummed expression
Eq.~(\ref{genres}). The general relation between the coefficients 
 $g_{0,n}$ and $\bar g_{0,n}$, is discussed in
Appendix~\ref{largenapp}, see in particular
Eq.~\eqref{eq:g0bardefapp}: it turns out (see
Tab.~\ref{tab:ggbar} and Ref.~\cite{MV})  that 
the known coefficients $g_{0,n}$ are rather
larger than $\bar g_{0,n}$. This is also the case for Drell-Yan
production~\cite{MV}. For this reason, at third order, where 
$g_{0,3}=\bar g_{0,3} +r_3$, with $r_3=114.7$, we will take 
$\bar g_{0,3}=0$ (rather than $g_{0,3}=0$)
as preferred choice (see also a corresponding discussion  in Ref.~\cite{MV}).
Coming now to the small $N$ expression Eq.~(\ref{Ch}), the
coefficients $c_{i,j}$ 
are collected  in
Appendix~\ref{smallnapp}, while explicit expressions for
$\gamma^{(i)}$ are given in Eq.~\eqref{eq:explicitgamma}. 
The
function $C^{(3)}_\text{approx}(N)$ is plotted in
Fig.~\ref{fig:approx_C3},
together with the soft approximation
(bounded by the two curves
$\Csoftm{1}^{(3)}(N)$ and $\Csoftm{2}^{(3)}(N)$) and the high energy
approximation  (given by $\Chard^{(3)}(N)$).

\begin{figure}[t]
 \centering
 \includegraphics[width=0.7\textwidth,page=1]{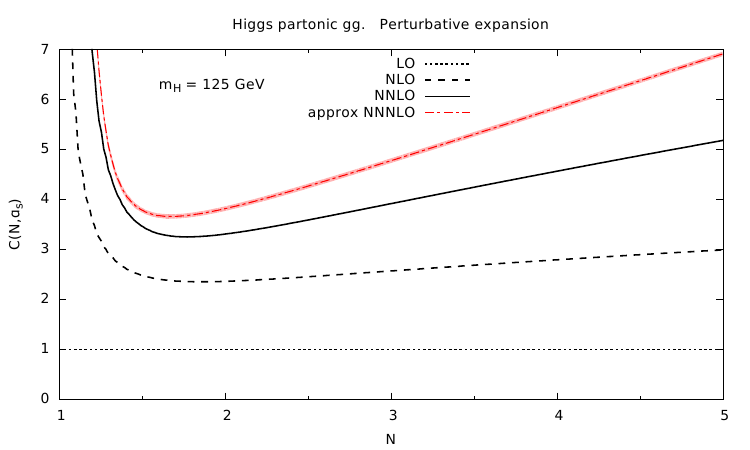}
 \caption{Behaviour of the perturbative expansion of $C(N,\as)$, with $\as(\mh^2)=0.1126$.}
 \label{fig:pert_expansion}
\end{figure}

We finally turn to the behaviour of the perturbative expansion of the
coefficient function.
In Fig.~\ref{fig:pert_expansion} we compare the NLO, NNLO and N$^3$LO
truncations of 
\beq \label{approxpertseries}
C_{\text{N$^3$LO}}(N,\as) = 1 + \as C^{(1)}(N) + \as^2 C^{(2)}(N) + \as^3 C_{\rm approx}^{(3)}(N).
\eeq
We note that  at moderately large $N\gtrsim 4$ (where
we expect our approximation to be very accurate) the 
$\Ord(\as^3)$ contribution is significant, so the convergence of the
series is quite slow. On the other hand, the saddle point argument of
Ref.~\cite{bfrsaddle} implies that the dominant contribution at LHC
energies comes from the region $N\sim 2$, where convergence is much
faster, though the N$^3$LO contribution
is still quite large.

\subsection{Hadron level results}
\label{sec:hl}

We now discuss the corresponding hadron level quantities.
To this purpose, we define the gluon channel $K$-factors
\beq\label{eq:Kfactor}
K_{gg}(\tau, \mh^2)=\frac{\sigma_{gg}(\tau,\mh^2)}{\sigma^{(0)}(\tau,\mh^2)}
=1+\as K_{gg}^{(1)}+\as^2 K_{gg}^{(2)}+\as^3 K_{gg}^{(3)}+\Ord(\as^4),
\eeq
where $\as=\as(\mh^2)$,
$\sigma_{gg}(\tau,\mh^2)$ is the contribution from the gluon
channel to the cross section Eq.~\eqref{eq:xs}, which  implies that
\beq
\sigma^{(0)}(\tau,\mh^2)=\tau \sigma_0(\mh^2,\as)\Lum_{gg}(\tau,\mh^2),
\eeq
and 
\beq\label{eq:Kn}
K_{ij}^{(n)}= \frac{1}{\Lum_{gg}(\tau,\mh^2)}
\int_\tau^1 \frac{dz}{z}\,\Lum_{ij}\(\frac{\tau}{z},\mh^2\)\,C^{(n)}_{ij}(z),
\eeq
and we use everywhere the NNLO expression of $\as$, and NNLO parton
distributions. 

We then compute the $K$-factors using various  approximations for the
coefficient function, and
compare them to each other and, at NLO and NNLO, to the exact
result. Specifically, besides our preferred approximation
Eq.~(\ref{approxfinal}) shown in
Figs.~\ref{fig:approx_coeff_funct}--\ref{fig:approx_C3}, we also show
results obtained using the soft contribution $\Csoft^{(n)}$ to the coefficient function
 (also shown in Figs.~\ref{fig:approx_coeff_funct}--\ref{fig:approx_C3}), as well as the $N$-soft
approximation Eq.~\eqref{eq:CresDlog} shown  in
 Fig.~\ref{fig:soft_comparison} and also determined with $\bar
 g_{0,3}=0$. This $N$-soft approximation is essentially the same as the
 N$^3$LO approximation previously published in Ref.~\cite{MV}, though
 here, unlike in Ref.~\cite{MV}, we include the full $\mt$ dependence.
We use the NNLO NNPDF2.1~\cite{NNPDF21} set of parton distribution function,
with $\as(\mz^2)=0.119$, and with the scale choice $\muf=\mur=\mh$. The
scale dependence will be studied in Sect.~\ref{sec:apphlhc} below.

\begin{figure}[t]
 \centering
 \includegraphics[width=0.49\textwidth,page=1]{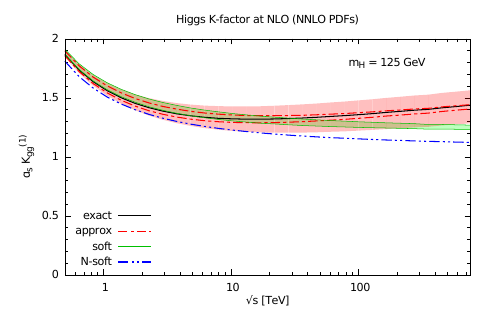}
 \includegraphics[width=0.49\textwidth,page=2]{paper_hadr_Kfact.pdf}
 \includegraphics[width=0.49\textwidth,page=3]{paper_hadr_Kfact.pdf}
 \caption{The NLO, NNLO and N$^3$LO contributions to the
   $K$-factor Eq.~\eqref{eq:Kn} in the gluon channel only  as a function of the
   collider energy $\sqrt{s}$, computed using the various
   approximations to the coefficient function shown in
   Figs.~\ref{fig:approx_coeff_funct}--\ref{fig:approx_C3}.  We also show
   approximation based on using the $N$-soft coefficient function of
   Fig.~\ref{fig:soft_comparison}, which at N$^3$LO is close to the result of Ref.~\cite{MV}.}
 \label{fig:hadronic_kfact}
\end{figure}
Results are shown in Fig.~\ref{fig:hadronic_kfact},
where the various contributions to the functions $K_{gg}^{(n)}$ are plotted as a
function of the collider energy $\sqrt{s}$. As the energy increases,
$\tau$ becomes smaller and one would expect small $z$ effects to
become more relevant. Indeed, we observe that the soft
approximations  deviates from the exact results,
while the full approximation reproduces well the shape of
the cross section for all $\sqrt{s}$.
On the other hand, at low energies the red and green curves (and
corresponding uncertainty bands)
tend to coincide, meaning that the small $z$ contribution has become negligible.
In that region, we also observe that the bottom edge of the
uncertainty band (which is obtained using $\Csoftm{2}$
Eq.~(\ref{eq:Csoft2})) better
approximates the exact result than the top (obtained using $\Csoftm{1}$
Eq.~(\ref{eq:Csoft1})).
Finally, we note the $N$-soft curve always undershoots
the exact result, the more so  at  higher
perturbative orders. This agrees with the behaviour of the
$N$-soft curve in
  Fig.~\ref{fig:soft_comparison}.

In Fig.~\ref{fig:hadronic_kfact} we also show our prediction for the N$^3$LO
$K$-factor. The third order
term  is quite large at small energy
$\sqrt{s}\sim1$~TeV, where  the soft contribution is dominant, but it
remains sizable even for $\sqrt{s}\sim10$~TeV, where it is almost half
of the NNLO. This slow convergence of the perturbative series may
make all order resummation mandatory.
In the low energy region the uncertainty band on our  soft approximation is
quite narrow, and the  $N$-soft curve is well below and outside it.
At very high energies the convergence of the perturbative expansion
seems to improve, but very slowly, and the uncertainty on our
prediction increases.


\section{N$^3$LO Higgs production at the  LHC}
\label{sec:apphlhc}

We now  concentrate on Higgs production at the LHC
at $\sqrt{s}=8$~TeV, with $\mh=125$~GeV.
In Table~\ref{table:fact} we
present the $K$-factor $K_{gg}$ Eq.~\eqref{eq:Kfactor}, computed 
again with the NNLO NNPDF2.1~\cite{NNPDF21} PDF set and
$\as(\mz^2)=0.119$.
Results at NLO, NNLO and N$^3$LO  are compared to the
exact results (when available) as well as the $N$-soft approximation
which, as mentioned in Sect.~\ref{sec:hl}, is essentially the same as
the approximation published in Ref.~\cite{MV}, from which it differs
because of the inclusion of finite top mass effects. We show results
for two choices of the renormalization scale, $\mur=\mh$ and
$\mur=\mh/2$, both with $\muf=\mh$: as we shall see below, the
factorization scale dependence is essentially negligible, even at LO.
The uncertainty  on our prediction has been determined as
discussed in Sect.~\ref{sec:partonic}.
Our approximate result agrees with the exact result at NLO and NNLO
within its stated uncertainty. The $N$-soft approximation 
leads to a systematically smaller result, the more so at higher perturbative orders.
\begin{table}[t]
\begin{center}
\begin{tabular}[c]{cccclccc}
  & \multicolumn{3}{c}{$\mur=\mh$} && \multicolumn{3}{c}{$\mur=\mh/2$} \\
\cline{2-4}\cline{6-8}
  & $C^{(n)}_\text{exact}$ & $C^{(n)}_\text{approx}$ & $\Cmv^{(n)}$ && $C^{(n)}_\text{exact}$ & $C^{(n)}_\text{approx}$ & $\Cmv^{(n)}$ \\
\midrule
 $\as   K_{gg}^{(1)}$ & $1.328$ & $1.330\pm0.099$ & $1.241$ && $1.262$ & $1.265\pm0.111$ & $1.167$ \\
 $\as^2 K_{gg}^{(2)}$ & $0.903$ & $0.968\pm0.088$ & $0.815$ && $0.795$ & $0.747\pm0.109$ & $0.558$ \\
 $\as^3 K_{gg}^{(3)}$ &   ---   & $0.495\pm0.051$ & $0.353$ &&   ---   & $0.279\pm0.070$ & $0.085$ \\
\end{tabular}
\end{center}
\caption{NLO, NNLO and N$^3$LO contributions to the gluon fusion
  $K$-factors Eq.~\eqref{eq:Kfactor}, computed with two different
  choices of the renormalization scale, $\muf=\mh$ and $\muf=\mh/2$.
  Our approximation is compared to
  the exact result (when available) and to the $N$-soft approximation,
  which up to NNLO coincides with the fixed-order truncation of the resummed
  result~\cite{Catani:2003zt,deFlorian:2012yg}, and at N$^3$LO is
  close to it and to the
  result of Ref.~\cite{MV} (see text). 
The uncertainty shown 
corresponds to the band in Fig.~\ref{fig:hadronic_kfact}.}
\label{table:fact}
\end{table}

We now turn to our result\footnote{In the published version of the paper
  the cross sections given in Eqs.~\eqref{eq:xsecmuR1},~\eqref{eq:xsecmuR0.5}
  are approximately $1\%$ larger due to a bug
  in the \texttt{ggHiggs} code, fixed in version 1.9.
  This shift is within  the stated uncertainty and none of our
  conclusions is affected.}
for the full N$^3$LO Higgs production cross section at central scale $\muf=\mh$
\begin{align}
\sigma_\text{approx}^{\text{N$^3$LO}}(\tau,\mh^2)&=\sigma^{(0)}(\tau,\mh^2)
\[\sum_{ij}\(\delta_{ig}\delta_{jg}+ \as K_{ij}^{(1)}+\as^2 K_{ij}^{(2)} \)+\as^3 K_{gg,\text{approx}}^{(3)}\]
\nonumber\\
&= \big( 22.41 \pm 0.32  +0.91\cdot 10^{-2} \bar g_{0,3} \big) \text{~pb} \qquad \text{for $\mur=\mh$}   \label{eq:xsecmuR1} \\
&= \big( 23.69 \pm 0.54  +1.55\cdot 10^{-2} \bar g_{0,3} \big) \text{~pb} \qquad \text{for $\mur=\mh/2$},\label{eq:xsecmuR0.5}
\end{align}
where the error  shown is our estimate of the uncertainty in our
approximation procedure, and we have separated off the contribution 
from the unknown coefficient $\bar g_{0,3}$, discussed in
Sect.~\ref{sec:partonicpl} above. As discussed in
Sect.~\ref{sec:partonic} our default choice is $\bar g_{0,3}=0$, on
the grounds that the perturbative behaviour of the $\bar g_{0,i}$
coefficients (see Table~\ref{tab:ggbar})
suggests that  $\bar g_{0,3}$ is possibly of order ten or
so (while the coefficient $g_{0,3}$ is likely to be rather larger,
perhaps of order hundred).

We have computed the LO, NLO and NNLO contributions to the cross
section, with full top mass
effects~\cite{spira2,harlander2}, and including 
all partonic subprocesses, while the
prediction at N$^3$LO only contains the approximate coefficient
function for the gluon channel.  We have cross-checked results in the
pointlike limit against the  \verb+ihixs+ code~\cite{ihixs}, and the
top mass dependence  against the numerical implementation of
Ref.~\cite{Bonciani:2007ex}.
   
With $\mur=\mh$, the N$^3$LO amounts to a $16\%$ correction  to the
NNLO prediction $\sigma_\text{NNLO}=19.33$~pb.
This correction is  larger than that found  in
Refs.~\cite{Catani:2003zt,deFlorian:2012yg} using  NNLL resummation,
which increases the NNLO result by  about $8\%$. If expanded out to
finite order, the resummed result of
Refs.~\cite{Catani:2003zt,deFlorian:2012yg} coincides with the
$N$-soft approximation, which at  N$^3$LO is by little more than
$1$~pb (corresponding to  6\%
of the NNLO) smaller (see Fig.~\ref{fig:hadronic_xs} below). 
Also,  the result of
Refs.~\cite{Catani:2003zt,deFlorian:2012yg} corresponds to taking
$g_{0,3}=0$ instead of  $\bar g_{0,3}=0$ as we do in our default
result, given that
in the
NNLL expression $g_0(\as)$, Eq.~\eqref{genres}, is included up to order
$\as^2$. With this choice, the   N$^3$LO is further reduced by about  5\%
of the NNLO, down to a correction of about 6\%. The extra 2\% or so in
Refs.~\cite{Catani:2003zt,deFlorian:2012yg} is accounted for by N$^4$LO
and higher orders. With $\mur=\mh/2$ as sometimes~\cite{ihixs}
advocated, the impact of the N$^3$LO corrections is reduced to $10\%$, but
the difference between our result and the $N$-soft
prediction (and thus also that based on NNLO resummation) 
increases, from about 5\% to about 7\%, see Fig.~\ref{fig:hadronic_xs} below.

\begin{figure}[t]
 \centering
 \includegraphics[width=0.49\textwidth,page=2]{paper_hadr_xsec.pdf}
 \includegraphics[width=0.49\textwidth,page=1]{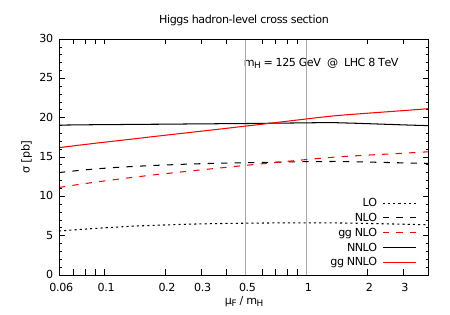}
 \caption{Dependence of the NLO and NNLO cross sections on
   the renormalization scale $\mur$ and factorization
   scale $\muf$. The curves labeled  $gg$ (N)NLO are obtained
   including all channels at (N)LO and the gluon-gluon
   contribution at (N)NLO. The two choices of renormalization scale
   used to compute Tab.~\ref{table:fact} are shown as vertical
   bars. The two corresponding choices of factorization scale are also
 shown (but only $\muf=\mh$ is used in Tab.~\ref{table:fact}).}
 \label{fig:hadronic_lower}
\end{figure}
We now study the dependence of the cross section on variations
of the renormalization and factorization scales, $\mur$ and $\muf$
respectively.  
We first show the scale dependence of the known NLO and NNLO
cross sections in Fig.~\ref{fig:hadronic_lower}, with the 
the choices of renormalization scale $\mur=\mh$ and $\mur=\mh/2$
   used to compute Tab.~\ref{table:fact} shown as vertical lines. The
   two choices  $\muf=\mh$ and $\muf=\mh/2$ are also shown, even
   though
only $\muf=\mh$ was used form Tab.~\ref{table:fact}.
We consider both
a simultaneous scale  variation in all partonic subprocesses (black
curves),  as  
well as the scale variation for the gluon-gluon subprocess only.
The renormalization scale dependence of the full result is not much
different from that of the gluon contribution.
The
factorization scale dependence instead is much stronger for the gluon
channel alone than for the full result. This cancellation of the factorization
scale dependence between partonic subchannels is a direct consequence of
the known structure of the Altarelli-Parisi equations. 
The factorization scale of the full result turns out to be essentially
negligible, thereby justifying the choice not to show the dependence
on it in Tab.~\ref{table:fact}.

The scale dependence of our   N$^3$LO result is
displayed in Fig.~\ref{fig:hadronic_xs}. 
We only show the renormalization scale dependence: 
the factorization scale dependence of the N$^3$LO result will be weaker than
that of the NNLO, which is already negligible. Also, our   N$^3$LO
result only includes the (dominant) gluon contribution, so its
factorization scale dependence would be misleadingly large, and
canceled by a  contribution from the quark channels.

\begin{figure}[t]
 \centering
 \includegraphics[width=0.7\textwidth,page=3]{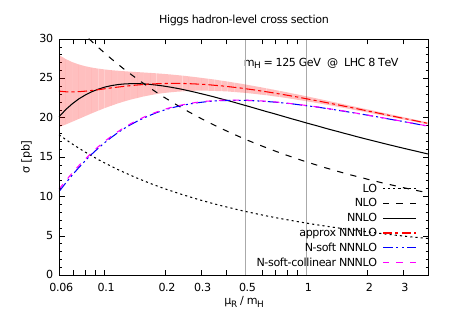}
 \caption{Dependence of the N$^3$LO cross section on
   the renormalization scale~$\mur$. The two choices of renormalization scale
   used to compute Tab.~\ref{table:fact} are shown as vertical
   bars. }
 \label{fig:hadronic_xs}
\end{figure}
The N$^3$LO contribution reduces the renormalization scale
dependence of the NNLO QCD result from $\pm10\% $ to $\pm  6\%$ if the
scale is varied in the range $0.5<\mur/\mh<2$. 
We also show the prediction obtained using the soft approximation
$\Cmv^{(3)}$, with $\bar g_{0,3}=0$,
i.e.\ essentially the approximation of Ref.~\cite{MV}, as well as the
prediction obtained by performing a collinear improvement of the
latter~\cite{Catani:2003zt} (labeled $N$-soft-collinear,
see Sect.~\ref{sec:largen}, Fig.~\ref{fig:soft_comparison}).
The fact that 
he $N$-soft result is rather smaller than our own is clearly seen. The
collinear improvement of Ref.~\cite{Catani:2003zt} has a negligible %
impact, and indeed it has therefore not been included~\cite{grazpriv}
in the recent phenomenological results of  Ref.~\cite{MV,deFlorian:2012yg}.
As seen in
Fig.~\ref{fig:hadronic_kfact}, for central scale choices $\mh/2\lesssim
\mur \lesssim \mh$ the 
the difference between our approximate result and the $N$-soft
approximation is  due almost entirely to our different way of
treating subleading soft terms, and this is thus the reason why correction is
more substantial than those of Refs.~\cite{MV,deFlorian:2012yg}
(note that in Ref.~\cite{MV} a smaller value of $\as(\mz)$ is
adopted,  which would lead
to a yet smaller result). The scale dependence of our result is
similar to that of the  $N$-soft result and its collinear improvement
(and thus to that of Refs.~\cite{MV,deFlorian:2012yg}) towards the
high end, but it has a different shape towards
lowers scales, where it is much weaker, partly due to the  matching 
with the small $N$ terms.

\section{Conclusions and Outlook} \label{sec:concl}

We have determined an approximate expression for the N$^3$LO Higgs
production cross section in gluon fusion, with finite top
mass. We have considered the dominant gluon channel only. 
Our approximation is based on combining information on the large $N$
and small $N$ singularities of the coefficient function, which are
determined from resummation, while making sure that they do not
interfere with each other, so that large $N$ terms do not introduce
spurious small $N$ singularities, and conversely.
Small $N$ resummation in unfortunately only known to the leading
logarithmic level (unlike large $N$ resummation, which is known up to
N$^3$LL order), so
our approximation looses accuracy at small $N$, but fortunately
Higgs production in gluon fusion is dominated by large $N$ terms down
to fairly moderate $N$ values~\cite{bfrsaddle}.

We
have found that at $\sqrt{s}=8$~TeV this correction leads to a $16 \%$
increase of the cross section for $\mur=\muf=\mh=125$~GeV and it noticeably
reduces the scale dependence of the NNLO result. Our correction is
larger than that previously found in Refs.~\cite{MV,deFlorian:2012yg}
essentially because of our different treatment of subleading soft
terms, while its scale dependence, especially towards lower scales, is
milder due to the matching to the small $N$ ``BFKL'' terms. The
difference becomes yet larger for lower scale choices.

The results presented here can be used to improve the prediction for
standard model Higgs production in  gluon fusion, 
and to the very least they provide an
estimate of the impact of higher order corrections on the currently
known NNLO result which is rather more reliable that the commonly used
scale variation.
A public code is available at
\begin{center}
\texttt{\href{http://www.ge.infn.it/~bonvini/higgs/}{http://www.ge.infn.it/$\sim$bonvini/higgs/}}
\end{center}

While we have  concentrated on the dominant  gluon channel
the inclusion of other partonic channels along the same lines is
possible. More interestingly, our approach could also be extended 
to the construction of approximate expressions for rapidity
distributions~\cite{Bonvini:2010tp,cfm}. Both are left for future
work, as well as the construction of a fully resummed result, in which
the large and small $N$ terms are included to all orders in
$\alpha_s$, and the extension to other processes, specifically
Drell-Yan production.

\section*{Acknowledgements}
We thank A.~Vicini and R.~Harlander for providing code which
implements the finite-mass terms of
Refs.~\cite{Bonciani:2007ex,harlander2} repectively.
SM wishes to thank Nigel Glover and Alex Mitov for interesting
discussions. Part of this work was done by RDB and SF during a visit 
to the Discovery Center of the Niels Bohr
Institute.
The work of SM is supported by UK's STFC. SF and GR are
supported by a PRIN2010 grant.

\subsection*{Note added}

After publication, we have found a bug in the \texttt{ggHiggs} code, fixed in version 1.9,
which affected our prediction at order $\as^3$.
As a consequence, Figs.~\ref{fig:approx_C3}, \ref{fig:pert_expansion}, \ref{fig:hadronic_kfact}
and \ref{fig:hadronic_xs}, last line of Tab.~\ref{table:fact} and Eqs.~\eqref{eq:xsecmuR1}, \eqref{eq:xsecmuR0.5}
have been updated.
The change is  within the uncertainty of the original result and therefore our
conclusions are unchanged.

\appendix
\allowdisplaybreaks
\section{Explicit results for the coefficients}
\label{sec:app_analytic}

\subsection{Large $N$ contributions}
\label{largenapp}

We present some results on Mellin transformation
of plus distributions which appear in perturbative calculations, defined by
\beq
\int_0^1 dz \, \plusq{f(z)} g(z) = \int_0^1 dz \, f(z) \[g(z)-g(1)\],
\eeq
where $g(z)$ is any test function, regular in $0\leq z\leq 1$.
The distributions
\begin{subequations}\label{eq:ddefs}
\begin{align}
\D_k(z)&\equiv \plus{\frac{\ln^k(1-z)}{1-z}} ,\label{ddef}\\
\Dm_k(z)&\equiv \plus{\frac{\ln^k\ln\frac1z}{\ln\frac1z}} ,\label{dmdef}\\
\Dh_k(z)&\equiv  \D_k(z)
+\[\frac{\ln^k\frac{1-z}{\sqrt{z}}}{1-z}-\frac{\ln^k(1-z)}{1-z}\],
\label{dhatdef}
\end{align}
\end{subequations}
can be obtained, respectively, as the $k$-th $\xi$-derivative of the 
generating distributions
\begin{subequations}\label{gendisdef}
\begin{align}
\D_k(z)&=\frac{d^k}{d\xi^k}  \plusq{(1-z)^{\xi-1}}\Bigg|_{\xi=0},\\
\Dm_k(z)&=\frac{d^k}{d\xi^k} \plus{\ln^{\xi-1}\frac1z}  \Bigg|_{\xi=0},\\
\Dh_k(z)&=\frac{d^k}{d\xi^k}  z^{-\xi/2}\plusq{(1-z)^{\xi-1}}\Bigg|_{\xi=0}.
\end{align}
\end{subequations}
The Mellin transforms 
\beq
\Mell[f]=\int_0^1dz\,z^{N-1}f(z)
\eeq
of the generating distributions Eq.~(\ref{gendisdef}) are easily computed:
\begin{subequations}\label{eq:Mellin_log_gen}
\begin{align} 
\Mell\[\plusq{(1-z)^{\xi-1}}\] &= \Gamma(\xi)\[\frac{\Gamma(N)}{\Gamma(N+\xi)} 
- \frac1{\Gamma(1+\xi)}\]\\
\Mell\[\plus{\ln^{\xi-1}\frac1z}\] &= \Gamma(\xi) \[N^{-\xi}-1\]
\label{eq:Mellin_log_gen2}\\
\Mell\[z^{-\xi/2}\plusq{(1-z)^{\xi-1}}\] &= 
  \Gamma(\xi)\[\frac{\Gamma(N-\xi/2)}{\Gamma(N+\xi/2)} 
- \frac1{\Gamma(1+\xi)}\]\label{eq:Mellin_log_gen4}
\end{align}
\end{subequations}
One finds~\cite{Bonvini:thesis}
\begin{subequations}\label{eq:Mellin_log}
\begin{align}\D_k(N)\equiv
\Mell\[\D_k(z)\]
&=\frac{1}{k+1}\sum_{j=0}^{k}\binom{k+1}{j}\,
\Gamma^{(j)}(1)\[ \Gamma(N)\,\Delta^{(k+1-j)}(N) - \Delta^{(k+1-j)}(1)\]
\label{eq:Mellin_log1}
\\ \Dm_k(N)\equiv
\Mell\[\Dm_k(z)\]
&=\frac{1}{k+1}\sum_{j=0}^k\binom{k+1}{j}\, \Gamma^{(j)}(1)\,
\ln^{k+1-j}\frac{1}{N}
\label{eq:Mellin_log2}
\\
\Dh_k(N)\equiv
\Mell\[\Dh_k(z)\]
&=\frac{1}{k+1}\sum_{j=0}^{k}\binom{k+1}{j}\,
\Gamma^{(j)}(1) \[\Upsilon^{(k+1-j)}(N,0)-\Delta^{(k+1-j)}(1) \]
\label{eq:Mellin_log4}
\end{align}
\end{subequations}
where we have defined
\begin{subequations}
\begin{align}
\Delta(\xi) &=\frac{1}{\Gamma(\xi) \label{eq:Delta_def}}\\
\Upsilon(N,\xi) &= \Gamma(N-\xi/2) \,\Delta(N+\xi/2)\label{eq:Upsilon}
\end{align}
\end{subequations}
and the superscripts in round brackets in $\Upsilon(N,\xi)$ denote
derivatives with respect to $\xi$.  We note that all the expressions
in Eqs.~\eqref{eq:Mellin_log} are of comparable complexity.  They all
share the same behavior at large $N$ term by term in the sums, which
implies that asymptotically they only differ by constant terms.
In particular,
\beq
\lim_{N\to\infty} \[\Dh_k(N)-\D_k(N)\] = 0,
\eeq
while the constants $d_k$ defined in Eq.~\eqref{eq:dk_def} are given by
\beq\label{eq:dk_def_app}
d_k \equiv \lim_{N\to\infty} \[\Dh_k(N)-\Dm_k(N)\]
= \frac1{k+1}\Gamma^{(k+1)}(1).
\eeq

The right-hand sides of Eqs.~\eqref{eq:Mellin_log} and \eqref{eq:dk_def_app}
are easily computed with the help of the recursion relations
\begin{subequations}
\begin{align}
\Gamma^{(k+1)}(N) &= \sum_{j=0}^{k} \binom{k}{j} \Gamma^{(k-j)}(N)\, \psi_j(N).
\\
\Delta^{(k+1)}(N) &= -\sum_{j=0}^{k} \binom{k}{j} \Delta^{(k-j)}(N)\, \psi_j(N)
\label{eq:Delta_rec}
\\
\Upsilon^{(k+1)}(N,0) &=
-\sum_{j=0}^k \binom{k}{j} 
\frac12 \[\frac{1}{2^j} + \frac{1}{(-2)^j} \]\, \Upsilon^{(k-j)}(N,0)\, 
\psi_j(N) 
\label{eq:Upsilon_rec}
\end{align}
\end{subequations}
In particular,  Eq.~\eqref{eq:Upsilon_rec} shows that
$\psi_j(N)$ with $j$ odd never appear in $\Upsilon^{(k)}(N,0)$.
More details can be found in Ref.~\cite{Bonvini:thesis}.

We now consider the coefficient function in the soft limit,
which is completely fixed by the coefficients $b_{n,k}$ and by the
function $g_0(\as)$ appearing in Eq.~\eqref{eq:CresDlog},
which we reproduce here:
\beq
C_{\rm res}(N,\as) = g_0(\as) \,\exp
\sum_{n=1}^\infty \as^n \sum_{k=0}^{n} b_{n,k}\,\Dm_k(N), 
\eeq
The coefficients $b_{n,k}$
depend only on soft gluon radiation, and therefore do not
depend on $\mh$ or $\mt$. They can be computed within the
effective theory in which the top is integrated out and the top loop
shrinks to a point (pointlike approximation).
On the other hand, $g_0(\as)$ depends on $\mh/\mt$, and
therefore it also depends on whether the pointlike approximation
is used or not. The dependence of $g_0(\as)$ on the ratio $\mh/\mt$
obviously affects logarithmic terms by interference in $C_{\rm res}(N,\as)$
but such dependence is under control.

We now list the explicit coefficients $b_{n,k}$ for $n=1,2,3$.
We omit the scale dependence, which can be restored by imposing scale invariance
of the hadronic cross section.
The order $\as$ coefficients are
\beq
b_{1,1} = \frac{4C_A}\pi, \qquad
b_{1,0} = 0.
\eeq
At order $\as^2$ we have
\begin{subequations}
\begin{align}
b_{2,2} &=\frac{1}{\pi^2}\(-\frac{11}3C_A^2+\frac23 C_An_f\)\\
b_{2,1} &=\frac{1}{\pi^2}
\[\(\frac{67}9-2\zeta_2\)C_A^2 - \frac{10}{9}C_An_f\]\\
b_{2,0} &=\frac{1}{\pi^2}\[ 
\(-\frac{101}{27}+\frac{11}3\zeta_2+\frac{7}2\zeta_3\)C_A^2
             +\(\frac{14}{27}-\frac23\zeta_2\)C_An_f\].
\end{align}
\end{subequations}
Finally, at order $\as^3$ we have~\cite{MV}
\begin{subequations}
\begin{align}
b_{3,3} &=\frac{1}{\pi^3}\[\frac{121}{27}C_A^3 
        -\frac{44}{27}C_A^2n_f+ \frac4{27} C_An_f^2\]\\
b_{3,2} &=\frac{1}{\pi^3}\[ \(-\frac{445}{27}+\frac{11}3\zeta_2\)C_A^3
       +\(\frac{289}{54}-\frac23\zeta_2\)C_A^2n_f
       +\frac12 C_AC_Fn_f  -\frac{10}{27}C_An_f^2\]\\
b_{3,1} &=\frac{1}{\pi^3}\bigg[ 
        \(\frac{15503}{648}-\frac{188}9\zeta_2-11\zeta_3+\frac{11}5\zeta_2^2\) C_A^3
        +\(-\frac{2051}{324}+6\zeta_2\) C_A^2n_f \nonumber\\
       &\qquad\quad+\(-\frac{55}{24}+2\zeta_3\) C_AC_Fn_f
        +\(\frac{25}{81}-\frac49\zeta_2\)C_An_f^2\bigg]\\
b_{3,0} &=\frac{1}{\pi^3}\bigg[ 
        \(-\frac{297029}{23328} + \frac{6139}{324}\zeta_2 +\frac{2509}{108}\zeta_3
        - \frac{187}{60}\zeta_2^2 -\frac{11}6 \zeta_2\zeta_3 -6\zeta_5\)C_A^3
      \nonumber\\ & \qquad\quad
        +\(\frac{31313}{11664}-\frac{1837}{324}\zeta_2
        -\frac{155}{36}\zeta_3+\frac{23}{30}\zeta_2^2\)C_A^2n_f
      \nonumber\\ & \qquad\quad
        +\(\frac{1711}{864}-\frac12\zeta_2-\frac{19}{18}\zeta_3-\frac15\zeta_2^2\)C_AC_Fn_f
        +\(-\frac{58}{729}+\frac{10}{27}\zeta_2+\frac5{27}\zeta_3\)C_An_f^2\bigg].
\end{align}
\end{subequations}

We now turn to the function
\beq
g_0(\as) = 1+\sum_{n=1}^\infty \as^n\, g_{0,n}.
\eeq
The first two terms of the expansions are known, and
for $\mh=125$~GeV, $\mt=172.5$~GeV and $n_f=5$, are given by%
\footnote{The NLO coefficient $g_{0,1}$ has been computed numerically
  using the implementation of Ref.~\cite{Bonciani:2007ex} of the exact
  result~\cite{spira2}.  The NNLO coefficient $g_{0,2}$ was computed in
  Ref.~\cite{harlander2} as an expansion in powers of $(\mh/\mt)^2$.
  We have checked that truncating the expansion to order 4 the result
  is accurate at the per mille level.  }
\beq
g_{0,1} = 8.7153;\qquad
g_{0,2} = 40.10,
\eeq
but $g_{0,3}$ is still unknown.
The function $\bar g_0(\as)$ is related to $g_0(\as)$ by
Eq.~\eqref{eq:g0bardef}, which we rewrite here using the explicit values of 
the $d_k$, Eq.~\eqref{eq:dk_def_app}:
\beq\label{eq:g0bardefapp}
\bar g_0(\as)=g_0(\as) 
\exp\[-\sum_{n=1}^\infty \as^n \sum_{k=0}^nb_{n,k}
\frac{\Gamma^{(k+1)}(1)}{k+1}\].
\eeq
We find
\beq
\bar g_0(\as) = 1+\sum_{n=1}^\infty \as^n \, \bar g_{0,n}, 
\eeq
where
\beq
\bar g_{0,n}=g_{0,n}-r_n
\eeq
and the $r_n$ can be read off Eq.~\eqref{eq:g0bardefapp}
order by order in $\as$. It is interesting to observe that
each $r_n$ depends on $g_{0,j}$ with $j<n$; in particular,
$r_3$ does not depend on the unknown coefficient $g_{0,3}$.
The numerical values of $r_n$ for $n=1,2,3$ are given in Table~\ref{tab:ggbar}.
\begin{table}[t]
  \begin{center}
  \begin{tabular}{cccc}
  $n$ & $\bar g_{0,n}$ & $r_n$ & $g_{0,n}$ \\
  \midrule
  $1$ & $4.9374$ & $3.7779$ & $8.7153$ \\
  $2$ & $10.92$  & $29.18$  & $40.10$  \\
  $3$ & unknown  & $114.7$  & unknown
  \end{tabular}
  \end{center}
  \caption{Numerical values of $g_0$ and $\bar g_0$ at various
    perturbative orders.}
  \label{tab:ggbar}
\end{table}
We note that $r_3$ is of order $10^2$, which is the order of magnitude
of a naive estimate of $g_{0,3}$ on the basis of the known values
of $g_{0,1},g_{0,2}$.

\subsection{Small $N$ contributions}
\label{smallnapp}

The coefficients $c_{i_1,i_2}$ of the small $N$ singularity, Eq.~\eqref{impact}, were expressed in terms of single and double integrals over the off-shell gluon virtualities in Refs~\cite{Higgsfinite,marzaniPhD}.
For $\mh=125$~GeV, $\mt=172.5$~GeV and $n_f=5$, their numerical values for $\muf=\mh$ are given in $\MSbar$ by
\begin{align}
  c_{1,0} &= 2.28 \nonumber\\
  c_{2,0} &= 4.12 &    c_{1,1} &= 5.66 \nonumber\\
  c_{3,0} &= 8.64 &    c_{2,1} &= 10.54
\end{align}
Factorization scale dependence can be easily restored by the substitutions
\begin{align}
  c_{1,0} &\to c_{1,0}+\lf &\lf = \ln\frac{\mh^2}{\muf^2} \nonumber\\
  c_{2,0} &\to c_{2,0}+c_{1,0}\lf + \frac{\lf^2}{2} \nonumber\\
  c_{1,1} &\to c_{1,1}+2c_{1,0}\lf+\lf^2 \nonumber\\
  c_{3,0} &\to c_{3,0}+c_{2,0}\lf + c_{1,0}\frac{\lf^2}2 + \frac{\lf^3}6 \nonumber\\
  c_{2,1} &\to c_{2,1}+(c_{2,0}+c_{1,1})\lf + 3c_{1,0}\frac{\lf^2}2 + \frac{\lf^3}2.
\end{align}

\end{document}